\documentclass[11pt]{article}
\usepackage[dvips]{graphicx}
\usepackage{float}
\usepackage{epsfig}
\usepackage{ulem}
\usepackage{latexsym,amsmath,amsfonts,amssymb}
\usepackage[latin1]{inputenc}
\usepackage{rotating}
\usepackage[american]{babel}
\usepackage[dvips]{graphicx}
\usepackage{bbm}
\usepackage{color}
\usepackage{slashed}
\usepackage[unicode]{hyperref}
\usepackage{lscape}
\usepackage{bigints}
\usepackage{enumerate}
\usepackage[shortlabels]{enumitem}
\usepackage{tikz}
\usetikzlibrary{decorations.pathreplacing}
\usetikzlibrary{shapes}
\usepackage{graphicx}
\usepackage{epsfig}
\usepackage{rotating}
\usepackage{amssymb}
\usepackage{subfigure}
\usepackage{dsfont}
\usepackage{psfrag}
\usepackage{amsmath,euscript,array,mathrsfs,amsfonts}
\usepackage{slashed}
\usepackage{array}
\usepackage{youngtab}
\usepackage{color}
\usepackage{bbold}

\usepackage{hyperref}
\hypersetup{
colorlinks = true,
linkcolor = red,
linktocpage = true,
citecolor = blue
}

\usepackage{tikz}
\usetikzlibrary{calc}
 \usetikzlibrary{decorations.text}
 \usetikzlibrary{shapes}

 \usetikzlibrary{decorations.pathmorphing}
\usetikzlibrary{decorations.pathreplacing}
\usetikzlibrary{arrows.meta}
\tikzset{
  >={To[length=5pt]}
  }
\usetikzlibrary{shapes, shapes.geometric, shapes.symbols, shapes.arrows, shapes.multipart, shapes.callouts, shapes.misc}
\tikzset{snake it/.style={decorate, decoration=snake}}
\tikzset{7brane/.style={circle, draw=black, fill=black,ultra thick,inner sep=1.5 pt, minimum size=1 pt,}, c/.default={4pt}}
\tikzset{cross/.style={cross out, draw=black,thick, minimum size=2*(#1-\pgflinewidth), inner sep=0pt, outer sep=0pt}, cross/.default={5pt}}
\tikzset{big7brane/.style={circle, draw=black, fill=black,ultra thick,inner sep=2.5 pt, minimum size=1 pt,}, c/.default={4pt}}
\tikzset{u/.style={circle, draw=black, fill=white,inner sep=2 pt, minimum size=2 pt,},f/.style={square, draw=black, fill=white,ultra thick,inner sep=4 pt, minimum size=2 pt,}}
\tikzset{so/.style={circle, draw=black, fill=red,inner sep=2 pt, minimum size=2 pt,},f/.style={square, draw=black, fill=white,ultra thick,inner sep=4 pt, minimum size=2 pt,}}
\tikzset{sp/.style={circle, draw=black, fill=blue,inner sep=2 pt, minimum size=2 pt,},f/.style={square, draw=black, fill=white,ultra thick,inner sep=4 pt, minimum size=2 pt,}}
\tikzset{uf/.style={rectangle, draw=black, fill=white,inner sep=3 pt, minimum size=4 pt,}}
\tikzset{spf/.style={rectangle, draw=black, fill=blue, thick,inner sep=3 pt, minimum size=4 pt, circle, draw=black, fill=blue,thick,inner sep=2 pt, minimum size=2 pt,},f/.style={square, draw=black, fill=white,ultra thick,inner sep=4 pt, minimum size=2 pt,}}
\tikzset{sof/.style={rectangle, draw=black, fill=red, thick,inner sep=3 pt, minimum size=4 pt,}}
\usetikzlibrary{positioning}
\usetikzlibrary{arrows}
\usetikzlibrary{decorations.pathreplacing}
\usetikzlibrary{shapes}

\makeatletter\def\l@subsubsection#1#2{}%
\makeatother

\renewcommand\theequation{\arabic{section}.\arabic{equation}} 

\usepackage{subfigure}

\def\cA{{\cal A}}

\def\cG{{\cal G}}

\def\CC{\ensuremath{\mathds C}}
\def\RR{\ensuremath{\mathds R}}
\def\ZZ{\ensuremath{\mathds Z}}

\DeclareMathOperator{\vol}{vol}

\DeclareMathOperator{\sech}{sech}
\DeclareMathOperator{\tr}{tr}

\DeclareMathOperator{\Li}{Li}
\DeclareMathOperator{\csch}{csch}

\def\Im{\mathop{\rm Im}}

\newcommand{\be}{\begin{equation}}
\newcommand{\ee}{\end{equation}}
\newcommand{\ba}{\begin{array}}
\newcommand{\ea}{\end{array}}

\def\im{Invent. Math.}

\def\hat{\widehat}
\def\a{\alpha}
\def\b{\beta}
\def\c{\gamma}
\def\d{\delta}
\def\f{\phi}               
\def\vf{\varphi}  
\def\tvf{\tilde{\varphi}}
\def\vp{\varphi}
\def\g{\gamma}
\def\h{\eta}
\def\j{\psi}
\def\k{\kappa}                    
\def\l{\lambda}
\def\m{\mu}
\def\n{\nu}
\def\o{\omega}  \def\w{\omega}

\def\q{\theta}  \def\th{\theta}                  
\def\r{\rho}                                     
\def\s{\sigma}                                   
\def\t{\tau}
\def\u{\upsilon}
\def\x{\xi}
\def\z{\zeta}
\def\pt{\tilde{\varphi}}
\def\tt{\tilde{\theta}}
\def\lab{\label}
\def\6{\partial}
\def\wg{\wedge}
\def\bpsi{\bar{\psi}}
\def\bt{\bar{\theta}}
\def\bvf{\bar{\varphi}}

\newcommand{\beq}{\begin{equation}}
\newcommand{\eeq}{\end{equation}}
\newcommand{\bea}{\begin{eqnarray}}
\newcommand{\eea}{\end{eqnarray}}

\newcommand{\beqs}{\begin{eqnarray}}
\newcommand{\eeqs}{\end{eqnarray}}
\newcommand{\bal}{\begin{aligned}}
\newcommand{\eal}{\end{aligned}}
\makeatletter
\newcommand\setItemnumber[1]{\setcounter{enum\romannumeral\@enumdepth}{\numexpr#1-1\relax}}
\makeatother
\begin{document}
\baselineskip=15.5pt
\pagestyle{plain}
\setcounter{page}{1}

\def\del{{\partial}}
\def\vev#1{\left\langle #1 \right\rangle}
\def\cn{{\cal N}}
\def\co{{\cal O}}


\def\IC{{\mathbb C}}
\def\IR{{\mathbb R}}
\def\IZ{{\mathbb Z}}
\def\RP{{\bf RP}}
\def\CP{{\bf CP}}
\def\Poincaré{{Poincar\'e }}
\def\tr{{\rm tr}}
\def\tp{{\tilde \Phi}}

\def\TL{\hfil$\displaystyle{##}$}
\def\TR{$\displaystyle{{}##}$\hfil}
\def\TC{\hfil$\displaystyle{##}$\hfil}
\def\TT{\hbox{##}}
\def\HLINE{\noalign{\vskip1\jot}\hline\noalign{\vskip1\jot}}
\def\seqalign#1#2{\vcenter{\openup1\jot
   \halign{\strut #1\cr #2 \cr}}}
\def\lbldef#1#2{\expandafter\gdef\csname #1\endcsname {#2}}
\def\eqn#1#2{\lbldef{#1}{(\ref{#1})}%
\begin{equation} #2 \label{#1} \end{equation}}
\def\eqalign#1{\vcenter{\openup1\jot
     \halign{\strut\span\TL & \span\TR\cr #1 \cr
    }}}

\def\eno#1{(\ref{#1})}
\def\href#1#2{#2}
\def\half{\frac{1}{2}}



\def\ads{{\it AdS}}
\def\adsp{{\it AdS}$_{p+2}$}
\def\cft{{\it CFT}}

\newcommand{\ber}{\begin{eqnarray}}
\newcommand{\eer}{\end{eqnarray}}

\newcommand{\beqar}{\begin{eqnarray}}
\newcommand{\cO}{{\cal O}}
\newcommand{\cT}{{\cal T}}
\newcommand{\cR}{{\cal R}}
\newcommand{\eeqar}{\end{eqnarray}}
\newcommand{\tht}{\thteta}
\newcommand{\lm}{\lambda}\newcommand{\Lm}{\Lambda}


\newcommand{\nonu}{\nonumber}
\newcommand{\oh}{\displaystyle{\frac{1}{2}}}
\newcommand{\dsl}
   {\kern.06em\hbox{\raise.15ex\hbox{$/$}\kern-.56em\hbox{$\partial$}}}
\newcommand{\as}{\not\!\! A}
\newcommand{\ps}{\not\! p}
\newcommand{\ks}{\not\! k}
\newcommand{\D}{{\cal{D}}}
\newcommand{\dv}{d^2x}
\newcommand{\Z}{{\cal Z}}
\newcommand{\N}{{\cal N}}
\newcommand{\Dsl}{\not\!\! D}
\newcommand{\Bsl}{\not\!\! B}
\newcommand{\Psl}{\not\!\! P}

\newcommand{\eeqarr}{\end{eqnarray}}


\def\del{{\delta^{\hbox{\sevenrm B}}}} \def\ex{{\hbox{\rm e}}}
\def\azb{A_{\bar z}} \def\az{A_z} \def\bzb{B_{\bar z}} \def\bz{B_z}
\def\czb{C_{\bar z}} \def\cz{C_z} \def\dzb{D_{\bar z}} \def\dz{D_z}
\def\im{{\hbox{\rm Im}}} \def\mod{{\hbox{\rm mod}}} \def\tr{{\hbox{\rm Tr}}}
\def\ch{{\hbox{\rm ch}}} \def\imp{{\hbox{\sevenrm Im}}}
\def\trp{{\hbox{\sevenrm Tr}}} \def\vol{{\hbox{\rm Vol}}}
\def\rl{\Lambda_{\hbox{\sevenrm R}}} \def\wl{\Lambda_{\hbox{\sevenrm W}}}
\def\fc{{\cal F}_{k+\cox}} \def\vev{vacuum expectation value}
\def\nodiv{\mid{\hbox{\hskip-7.8pt/}}}
\def\ie{{\em i.e.}}
\def\ie{\hbox{\it i.e.}}

\def\CC{{\mathchoice
{\rm C\mkern-8mu\vrule height1.45ex depth-.05ex
width.05em\mkern9mu\kern-.05em}
{\rm C\mkern-8mu\vrule height1.45ex depth-.05ex
width.05em\mkern9mu\kern-.05em}
{\rm C\mkern-8mu\vrule height1ex depth-.07ex
width.035em\mkern9mu\kern-.035em}
{\rm C\mkern-8mu\vrule height.65ex depth-.1ex
width.025em\mkern8mu\kern-.025em}}}

\def\RR{{\rm I\kern-1.6pt {\rm R}}}
\def\NN{{\rm I\!N}}
\def\ZZ{{\rm Z}\kern-3.8pt {\rm Z} \kern2pt}
\def\IB{\relax{\rm I\kern-.18em B}}
\def\ID{\relax{\rm I\kern-.18em D}}
\def\II{\relax{\rm I\kern-.18em I}}
\def\IP{\relax{\rm I\kern-.18em P}}
\newcommand{\CS}{{\scriptstyle {\rm CS}}}
\newcommand{\CSs}{{\scriptscriptstyle {\rm CS}}}
\newcommand{\rc}{\nonumber\\}
\newcommand{\bear}{\begin{eqnarray}}
\newcommand{\eear}{\end{eqnarray}}

\newcommand{\LL}{{\cal L}}

\def\mani{{\cal M}}
\def\calo{{\cal O}}
\def\calb{{\cal B}}
\def\calw{{\cal W}}
\def\calz{{\cal Z}}
\def\cald{{\cal D}}
\def\calc{{\cal C}}

\def\to{\rightarrow}
\def\ele{{\hbox{\sevenrm L}}}
\def\ere{{\hbox{\sevenrm R}}}
\def\zb{{\bar z}}
\def\wb{{\bar w}}
\def\nodiv{\mid{\hbox{\hskip-7.8pt/}}}
\def\menos{\hbox{\hskip-2.9pt}}
\def\dr{\dot R_}
\def\drr{\dot r_}
\def\ds{\dot s_}
\def\da{\dot A_}
\def\dga{\dot \gamma_}
\def\ga{\gamma_}
\def\dal{\dot\alpha_}
\def\al{\alpha_}
\def\cl{{closed}}
\def\cls{{closing}}
\def\vev{vacuum expectation value}
\def\tr{{\rm Tr}}
\def\to{\rightarrow}
\def\too{\longrightarrow}


\def\a{\alpha}
\def\b{\beta}
\def\c{\gamma}
\def\d{\delta}
\def\e{\epsilon}           
\def\F{\Phi}
\def\f{\phi}               
\def\vf{\varphi}  \def\tvf{\tilde{\varphi}}
\def\vp{\varphi}
\def\g{\gamma}
\def\h{\eta}
\def\j{\psi}
\def\k{\kappa}                    
\def\l{\lambda}
\def\m{\mu}
\def\n{\nu}
\def\o{\omega}  \def\w{\omega}
\def\q{\theta}  \def\th{\theta}                  
\def\r{\rho}                                     
\def\s{\sigma}                                   
\def\t{\tau}
\def\u{\upsilon}
\def\x{\xi}
\def\X{\Xi}
\def\z{\zeta}
\def\pt{\tilde{\varphi}}
\def\tt{\tilde{\theta}}
\def\lab{\label}
\def\6{\partial}
\def\wg{\wedge}
\def\atanh{{\rm arctanh}}
\def\bpsi{\bar{\psi}}
\def\bt{\bar{\theta}}
\def\bvf{\bar{\varphi}}

%



\newfont{\namefont}{cmr10}
\newfont{\addfont}{cmti7 scaled 1440}
\newfont{\boldmathfont}{cmbx10}
\newfont{\headfontb}{cmbx10 scaled 1728}





\newcommand{\re}{\,\mathbb{R}\mbox{e}\,}
\newcommand{\hyph}[1]{$#1$\nobreakdash-\hspace{0pt}}
\providecommand{\abs}[1]{\lvert#1\rvert}
\newcommand{\Nugual}[1]{$\mathcal{N}= #1 $}
\newcommand{\sub}[2]{#1_\text{#2}}
\newcommand{\partfrac}[2]{\frac{\partial #1}{\partial #2}}
\newcommand{\bsp}[1]{\begin{equation} \begin{split} #1 \end{split} \end{equation}}
\newcommand{\calF}{\mathcal{F}}
\newcommand{\calO}{\mathcal{O}}
\newcommand{\calM}{\mathcal{M}}
\newcommand{\calV}{\mathcal{V}}
\newcommand{\bbZ}{\mathbb{Z}}
\newcommand{\bbC}{\mathbb{C}}
\newcommand{\cK}{{\cal K}}

\newcommand{\Thq}{\Theta\left(\r-\r_q\right)}
\newcommand{\Dq}{\d\left(\r-\r_q\right)}
\newcommand{\kten}{\kappa^2_{\left(10\right)}}
\newcommand{\pbi}[1]{\imath^*\left(#1\right)}
\newcommand{\ho}{\hat{\omega}}
\newcommand{\tth}{\tilde{\th}}
\newcommand{\tf}{\tilde{\f}}
\newcommand{\tj}{\tilde{\j}}
\newcommand{\tw}{\tilde{\omega}}
\newcommand{\tz}{\tilde{z}}
\newcommand{\prj}[2]{(\partial_r{#1})(\partial_{\j}{#2})-(\partial_r{#2})(\partial_{\j}{#1})}
\def\atanh{{\rm arctanh}}
\def\sech{{\rm sech}}
\def\csch{{\rm csch}}
\allowdisplaybreaks[1]

\def\red{\textcolor[rgb]{0.98,0.00,0.00}}

\newcommand{\Dan}[1] {{\textcolor{blue}{#1}}}

\numberwithin{equation}{section}

\newcommand{\Tr}{\mbox{Tr}}    


%

\setcounter{footnote}{0}
\renewcommand{\theequation}{{\rm\thesection.\arabic{equation}}}

\begin{titlepage}

\begin{center}

\vskip .5in 
\noindent

{\Large \bf{ Wilson loops for 5d and 3d conformal linear quivers} }
\bigskip\medskip

Ali Fatemiabhari \footnote{a.fatemiabhari.2127756@swansea.ac.uk} and Carlos Nunez\footnote{c.nunez@swansea.ac.uk}\\

\bigskip\medskip
{\small 
Department of Physics, Swansea University, Swansea SA2 8PP, United Kingdom}

\vskip .5cm 
\vskip .9cm 
     	{\bf Abstract }\vskip .1in
\end{center}

\noindent
Within the electrostatic formulation of holographic duals to (balanced) conformal quivers in five and  three dimensions, we study the expressions for Wilson loops in antisymmetric representations. We derive general expressions for various quantities participating in the formalism (VEV of Wilson loops, representation, gauge-node) and apply these to examples, connecting some results present in the bibliography. In the case of three dimensional quivers, we present a relation between Wilson loops in an 'electric' and in the 'magnetic/mirror' descriptions. In a very detailed appendix, we relate the electrostatic and holomorphic description of the holographic duals to these SCFTS.
 
 \noindent
\vskip .5cm
\vskip .5cm
\vfill
\eject

\end{titlepage}

\setcounter{footnote}{0}

\small{
\tableofcontents}

\normalsize

\newpage
\renewcommand{\theequation}{{\rm\thesection.\arabic{equation}}}
\section{Introduction}

The Maldacena conjecture, or AdS/CFT \cite{Maldacena:1997re} motivates the study of both gravity and field theory  topics. In particular, the study of supersymmetric and conformal field theories in diverse dimensions. 
In relation to this, various efforts have been dedicated to  the classification of Type II or M-theory backgrounds with AdS$_{d+1}$ factors.
These backgrounds are proposed as holographic duals  to (encoding semi-classically the highly quantum dynamics of) SCFTs in $d$ dimensions with different amounts of SUSY. For the case in which the solutions are half-maximally supersymmetric, important progress in classifying string backgrounds and the mapping to families of quantum field theories has been achieved. 

A lot of work has been done along the lines described above. In this paper, we focus our attention on the case of conformal and supersymmetric linear quiver field theories in three and five dimensions preserving eight Poincare supercharges. This is the framework in which  this paper should be read.

In the case of three dimensional ${\cal N}=4$ SCFTs, the field theoretical aspects of linear quivers  presented in \cite{Gaiotto:2008ak} were discussed holographically in \cite{DHoker:2007hhe}-\cite{Akhond:2021ffz} among other works.
The case of ${\cal N}=1$ five dimensional linear quiver SCFTs (with eight Poincar\'e supercharges) was initially analysed holographically  in \cite{DHoker:2016ujz}. 
A non-exhaustive list of papers testing  the correspondence and analysing predictions derived for this case are \cite{DHoker:2016ujz}-\cite{Legramandi:2021uds}. 

In this work, we are mainly interested on Wilson loops.
These gauge invariant observables are of outmost importance and have been
profusely studied in the context of AdS/CFT. See \cite{Maldacena:1998im}-\cite{Drukker:2019bev}, for a  brief collection of papers on the topic.  
In SUSY gauge theories, the Wilson loop is a particularly interesting observable, as it can be computed exactly. Their relevance to AdS/CFT is the addition they make to the already rich dictionary between gauge theory and string theory. Studies in four dimensional ${\cal N}=4$ SYM have been done for over twenty years. Less understood is the case of SUSY Wilson loops in three or five dimensional SCFTs. This paper focuses on this particular problem.

In three dimensions, the Wilson loop in ${\cal N}=4$ supersymmetric field theories is labelled by a representation $\mathbb{R}$ of a given gauge group,
\begin{equation}
W_{\mathbb{R}}= \Tr_{\mathbb{R}} {\cal P} e^{i\oint \left(A_\mu \dot{x}^\mu + \sigma_3 \sqrt{-\dot{x}^2} \right) d\tau}  .\label{WLgenerico}
\end{equation}
Where $\sigma_3$
is one of the three scalars in the vector multiplet. The original  bosonic symmetries of the SCFT $SO(2,3)\times SU(2)_L\times SU(2)_R$ is broken by the presence of the operator in eq.(\ref{WLgenerico}) into $SU(1,2)\times U(1)\times SU(2)_L\times U(1)_R$. This is the algebra of superconformal quantum mechanics. At low energies, when the three dimensional QFT reaches a fixed point, the Wilson loop becomes a conformal line operator. For further studies on three dimensional SCFTs and their Wilson loops see \cite{Assel:2015oxa}-\cite{Drukker:2019bev}.

The situation in five dimensional SCFTs is similar. The Wilson loop is given by eq.(\ref{WLgenerico}), where $\sigma_3$  is in this case the adjoint scalar in the vector multiplet. The Wilson loop in five dimensional SCFTs preserves the $SU(2)_R$ of the theory and breaks $SO(2,5)\to SU(1,1)\times SO(4)$. For further studies on Wilson loops in five dimensional SCFTs see \cite{Assel:2018rcw},

In this paper we rely on the calculations with  Wilson loops described in \cite{Uhlemann:2020bek},\cite{Coccia:2021lpp}.
 We use the electrostatic formalism described in \cite{Akhond:2021ffz},  \cite{Legramandi:2021uds}, translating the results of  \cite{Uhlemann:2020bek},  \cite{Coccia:2021lpp}  into the
the electrostatic formulation. 
An  advantage of the formalism presented here is that some other calculations and the interpretation of the solutions may be easier to perform using our electrostatic viewpoint. Also, the electrostatic formalism makes clear certain analogies between systems in different dimensions. We discuss  mirror symmetry in the three dimensional case, proposing a relation between Wilson Loops in both mirror descriptions.

The organisation of the material in this paper is the following: in Section \ref{sectiongeometry} we summarise the electrostatic formalism to construct holographic duals to  balanced-quiver SCFTs  in dimensions five and three, with emphasis on the analogies between these two cases. The general characteristics described in this section extend to SCFTs in 1,2,4 and 6 dimensions.
In Section \ref{wilson5-3} we summarise the result of the works \cite{Uhlemann:2020bek},  \cite{Coccia:2021lpp} in the electrostatic language, relegating to an appendix the careful derivation of these results. We discuss two examples in full detail, clarifying  and connecting different results in the bibliography.
In Section \ref{mirrorsection}
we discuss aspects  of Mirror symmetry, as seen by the electrostatic formalism. In particular, we derive an expression relating the Wilson loops in a given representation in both electric and magnetic description.  In Section \ref{concl}, we  summarise and close this paper, proposing some topics for further study.

In the appendixes, we briefly elaborate on the matrix model perspective of our results and we describe precisely the translation
between the 'holomorphic' formalism of \cite{DHoker:2007hhe} and the electrostatic perspective pushed in this paper, making clear the connection with S-duality.

\section{Supergravity backgrounds}\label{sectiongeometry}
In this section, we discuss  the supergravity solutions used in this paper. 
We summarise the backgrounds preserving eight Poincare supercharges (${\cal N}=1$ SUSY in five dimensions and ${\cal N}=4$ in three dimensions). 
Supersymmetry is preserved subject to a  linear PDE being satisfied. We solve the PDE and briefly comment on the quantised charges and the associated  dual CFTs.
\subsection{The Type IIB Backgrounds dual to 5d SCFTs}
We present an infinite family of Type IIB backgrounds  preserving eight Poincar\'e supersymmetries with an AdS$_6$ factor. The space also contains a two sphere parameterised by  coordinates $(\theta,\varphi)$. The isometries of this manifold correspond with the $SO(2,5)\times SU(2)_R$  bosonic global symmetry of the dual ${\cal N}=1$ five dimensional SCFTs.

The full configuration consists of a metric, dilaton, $B_2$-field in the NS sector and $C_2$ and $C_0$ fields in the Ramond sector. The configuration is written in terms of a potential function $V_5(\sigma,\eta)$ that solves a linear partial differential equation written below.
The type IIB background in string frame is \cite{Legramandi:2021uds}, 
\begin{eqnarray}
& & ds_{10,st}^2= f_1(\sigma,\eta)\Big[ds^2(\text{AdS}_6) + f_2(\sigma,\eta)ds^2(S^2) + f_3(\sigma,\eta)(d\sigma^2+d\eta^2) \Big],\;\;e^{-2\Phi}=f_6(\sigma,\eta) , \nonumber\\[2mm]
& & B_2=f_4(\sigma,\eta) \text{Vol}(S^2),\;\;C_2= f_5(\sigma,\eta) \text{Vol}(S^2),\;\;\; C_0= f_7(\sigma,\eta), \label{backgroundrescaled}\\[2mm]
& & f_1= \frac{3 \pi}{2}\sqrt{\sigma^2 +\frac{3\sigma \partial_\sigma V_5}{\partial^2_\eta V_5}},\;\; f_2= \frac{\partial_\sigma V_5 \partial^2_\eta V_5}{3\Lambda},\;\;f_3= \frac{\partial^2_\eta V_5}{3\sigma \partial_\sigma V_5},\;\;\Lambda=\sigma(\partial_\sigma\partial_\eta V_5)^2 + (\partial_\sigma V_5-\sigma \partial^2_\sigma V_5)  \partial^2_\eta V_5,\nonumber\\[2mm]
& & f_4=\frac{\pi}{2}\left(\eta -\frac{(\sigma \partial_\sigma V_5) (\partial_\sigma\partial_\eta V_5)}{\Lambda} \right),\;\;\;\; f_5=\pi\left( V_5 - \frac{\sigma\partial_\sigma V_5}{\Lambda} (\partial_\eta V_5 (\partial_\sigma \partial_\eta V_5) -3 (\partial^2_\eta V_5)(\partial_\sigma V_5)) \right),\nonumber\\[2mm]
& & f_6=12\frac{\sigma^2 \partial_\sigma V_5 \partial^2_\eta V_5}{(3 \partial_\sigma V_5 +\sigma \partial^2_\eta V_5)^2}\Lambda,\;\;\;\; f_7=2\left( \partial_\eta V_5 + \frac{(3\sigma \partial_\sigma V_5) (\partial_\sigma\partial_\eta V_5 )}{3\partial_\sigma V_5 +\sigma \partial^2_\eta V_5}  \right).\nonumber
\end{eqnarray}
The function $V_5(\sigma,\eta)$ solves
\begin{equation}
\partial_\sigma \left(\sigma^2 \partial_\sigma V_5\right) +\sigma^2 \partial^2_\eta V_5=0.\label{diffeq5} \end{equation}
The paper \cite{Legramandi:2021uds} proves that this infinite family of backgrounds is 
in exact correspondence with the solutions discussed in  \cite{DHoker:2016ujz}-\cite{Gutperle:2018vdd}.

Let us briefly summarise the  study  of \cite{Legramandi:2021uds} for the  PDE, with boundary conditions leading  to a proper interpretation of the solutions, with quantised Page charges  and avoiding badly-singular behaviours.

\subsubsection{Resolution of the PDE and quantisation of charges}\label{resPDE}
We make the change
$
V_5(\sigma,\eta)=\frac{\hat{V}_5 (\sigma,\eta)}{\sigma},$
which implies that the PDE in (\ref{diffeq5}) reads like a Laplace equation in flat space,
\begin{equation}
\partial^2_\sigma \hat{V}_5 + \partial_\eta^2 \hat{V}_5=0.\label{eqfinal}
\end{equation}
We choose the variable $\eta$ to be bounded in the interval $[0,P]$ and $\sigma$ to range over the real axis $-\infty<\sigma<\infty$. We impose the boundary conditions,
\begin{eqnarray}
& & \hat{V}_5(\sigma\to\pm\infty,\eta)=0,\;\;\;\;\;\hat{V}_5(\sigma, \eta=0)= \hat{V}_5(\sigma, \eta=P)=0.\nonumber\\
& & \lim_{\epsilon\to 0}\left(\partial_\sigma \hat{V}_5(\sigma=+\epsilon,\eta)- \partial_\sigma \hat{V}_5(\sigma=-\epsilon,\eta)\right)= {\cal R}(\eta).\label{bc}
\end{eqnarray}

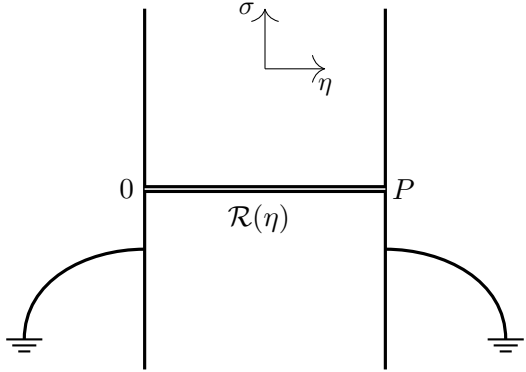
\begin{figure}
	\centering
	\begin{tikzpicture}[scale=0.8]
	\draw[very thick] (0,0) -- (0,6); 
	\draw[very thick] (4,0) -- (4,6);
	\draw[very thick] (0,2) to [out=-180,in=90] (-2,0.5);
	\draw[very thick] (4,2) to [out=0,in=90] (6,0.5);
	\draw[thick] (5.7,0.5) -- (6.3,0.5);
	\draw[thick] (5.8,0.4) -- (6.2,0.4);
	\draw[thick] (5.9,0.3) -- (6.1,0.3);
	\draw[thick] (-1.7,0.5) -- (-2.3,0.5);
	\draw[thick] (-1.8,0.4) -- (-2.2,0.4);
	\draw[thick] (-1.9,0.3) -- (-2.1,0.3);
	\draw[->] (2,5) -- (2,6);
	\draw[->] (2,5) -- (3,5);
	\node at (1.7,6) {\small $\sigma$};
	\node at (3,4.7) {\small $\eta$};
	\draw[very thick,double] (0,3) -- (4,3); 
	\node at (-0.3,3) {$0$};
	\node at (4.3,3) {$P$};
	\node at (1.9,2.5) {$\mathcal{R}(\eta)$};
	\end{tikzpicture}
	\caption{Depiction of the electrostatic problem for $\hat{V}_5$. The two conducting planes at $\eta=0,P$ have zero potential, while at $\sigma=0$ we have a charge distribution equal to $\cal R (\eta)$.}
	\label{fig:elect_problem}
\end{figure}
These can be interpreted as the boundary conditions for the electrostatic problem of two conducting planes (at zero electrostatic potential) as depicted in Figure \ref{fig:elect_problem}. The conducting planes extend over the $\sigma$-direction and are placed at $\eta=0$ and $\eta=P$. We also have a charge density ${\cal R}(\eta)$ at $\sigma=0$, extended along $0\leq \eta\leq P$, as indicated by the difference of the normal components of the electric field. 
The function ${\cal R}(\eta)$ can be taken to satisfy  
\begin{equation}
{\cal R}(\eta=0)={\cal R} (\eta=P)=0.\label{condicion-f}
\end{equation}
We refer to this in eq.(\ref{condicion-f}) as a situation {\it without offsets}.  Otherwise, if ${\cal R}(\eta)$ is non-zero at  either $\eta=0$ or $\eta=P$ we refer to as a situation  {\it with offsets}.

The solution is found by separating variables; see \cite{Legramandi:2021uds} for the details.  It is convenient to Fourier expand the function ${\cal R}(\eta)$ as,
\begin{equation}
{\cal R}(\eta)= \sum_{k=1}^\infty \mathcal{R}_k \sin \left(\frac{k\pi}{P}\eta \right) ,\;\;\;\; \mathcal{R}_k=\frac{2}{P} \int_0^P R(\eta) \sin\left( \frac{k \pi \eta}{P}\right) d\eta.\label{rankfunction}
\end{equation}
Following  \cite{Legramandi:2021uds},
the solution reads,
\begin{equation}
\label{eq:fourier_vhat} 
\hat{V}_5(\sigma,\eta)= \sum_{k=1}^\infty a_k \sin\left(\frac{k\pi}{P}\eta \right) {e^{-\frac{k\pi}{P}|\sigma|}},\;\;\;\;a_k= \frac{P}{2 \pi k } \mathcal{R}_k.
\end{equation}
Notice that we can introduce a complex variable
\begin{equation}
z= \sigma - i\eta,\nonumber
\end{equation}
and write the potential $\hat{V}_5=\sigma V_5$ as a harmonic function for both $\sigma > 0$ and $\sigma < 0$
\begin{equation}
\hat{V}_5(\sigma,\eta)= \begin{cases}
\sum_{k=1}^\infty  \frac{a_k}{2 i} \left( e^{-\frac{k\pi}{P}z} - e^{-\frac{k \pi}{{P}} \bar{z} }\right) & \sigma \ge 0, \\[2mm]
\sum_{k=1}^\infty  \frac{i a_k}{2 } \left( e^{\frac{k\pi}{P} z} - e^{\frac{k \pi}{{P}} \bar{z} }\right) & \sigma < 0 .
\end{cases}\label{potencialestext}
\end{equation}
$\hat{V}_5$ can therefore be expressed as the real part of a holomorphic function, and regularity is broken at $\sigma = 0$ due to the charge density in the electrostatic problem. See \cite{Legramandi:2021uds} and Appendix \ref{appendixmapping}, for translation between our formalism and the holomorphic one in \cite{DHoker:2016ujz,DHoker:2016ysh}.

The reader can check that the potentials in eqs.(\ref{eq:fourier_vhat})-(\ref{potencialestext}) solve the equations (\ref{diffeq5}),(\ref{eqfinal}) subject to the conditions in eq.(\ref{bc}). 

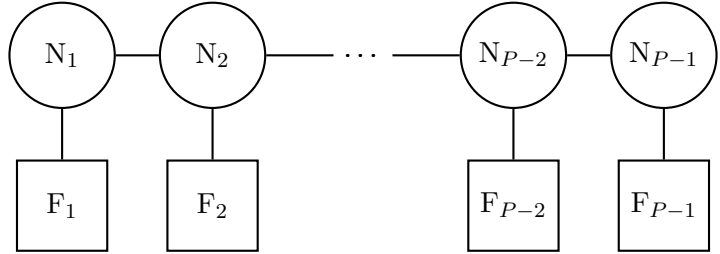
\begin{figure}[h!]
\begin{center}
	\begin{tikzpicture}
	\node (1) at (-4,0) [circle,draw,thick,minimum size=1.4cm] {N$_1$};
	\node (2) at (-2,0) [circle,draw,thick,minimum size=1.4cm] {N$_2$};
	\node (3) at (0,0)  {$\dots$};
	\node (5) at (4,0) [circle,draw,thick,minimum size=1.4cm] {N$_{P-1}$};
	\node (4) at (2,0) [circle,draw,thick,minimum size=1.4cm] {N$_{P-2}$};
	\draw[thick] (1) -- (2) -- (3) -- (4) -- (5);
	\node (1b) at (-4,-2) [rectangle,draw,thick,minimum size=1.2cm] {F$_1$};
	\node (2b) at (-2,-2) [rectangle,draw,thick,minimum size=1.2cm] {F$_2$};
	\node (3b) at (0,0)  {$\dots$};
	\node (5b) at (4,-2) [rectangle,draw,thick,minimum size=1.2cm] {F$_{P-1}$};
	\node (4b) at (2,-2) [rectangle,draw,thick,minimum size=1.2cm] {F$_{P-2}$};
	\draw[thick] (1) -- (1b);
	\draw[thick] (2) -- (2b);
	\draw[thick] (4) -- (4b);
	\draw[thick] (5) -- (5b);
	\end{tikzpicture}
\end{center}
\caption{Long quiver of length $P-1$ with gauge nodes $N_i$ and flavour nodes $F_i$. The quiver is \textit{balanced} if $F_i = 2 N_i - N_{i-1}-N_{i+1}$.}
\label{fig:quiverfiga}
\end{figure}

Imposing the quantisation of the conserved  Page charges in eq.(\ref{backgroundrescaled}), the authors of 
 \cite{Legramandi:2021uds} found that the function ${\cal R}(\eta)$ must be a convex piecewise linear function. 
\begin{equation}
 {\cal R}(\eta) = \begin{cases} 
       N_0+ ( N_1-N_0) \eta & 0\leq \eta \leq 1 \\
         N_l+ (N_{l+1} - N_l)(\eta-l) & l \leq \eta\leq l+1,\;\;\; l:=1,...., P-2\\\label{rankfull}
 %
         N_P+ (N_{P-1} -N_P)(P-\eta) & (P-1)\leq \eta\leq P . 
      \end{cases}
\end{equation}
For $N_0=N_P=0$ this is a rank function without off-sets. Otherwise, it has off-sets.
 In the case of no-offsets, the values of the quantised brane charges in each interval $[k, k+1]$ and in the system as a whole have been computed in  \cite{Legramandi:2021uds},
\begin{eqnarray}
& & Q_{NS5,total} = P \, \label{chargesfinal}\\
& &  Q_{D7}[k, k+1]= {\cal R}''(k)=(2 N_{k} - N_{k+1}- N_{k-1}),\; Q_{D7,total}=(N_1+ N_{P-1})= \int_0^P {\cal R}'' (\eta) d \eta ,\nonumber\\
 & & Q_{D5}[k,k+1] = {\cal R}(\eta) -{\cal R}'(\eta) (\eta- \Delta) =N_k\, ,\;\;\;\;\; Q_{D5,total}=\int_0^P {\cal R} ~d\eta.\nonumber 
\end{eqnarray}
For the generic rank function ${\cal R} (\eta)$ quoted in eq.(\ref{rankfull}), the supergravity background is proposed to be dual to the strongly coupled, UV-fixed point of the 
quiver in Figure \ref{fig:quiverfiga} for which 
$F_i= 2 N_i- N_{i+1} -N_{i-1}$.  In other words, the quiver is balanced.

We now discuss the Type IIB backgrounds dual to three dimensional  SCFTs  preserving eight SUSYs. The formalism is very much analogous to the five dimensional one, hence we will be more sketchy. All the details can be found in 
\cite{Akhond:2021ffz}.
\subsection{The Type IIB backgrounds dual to 3d SCFTs}\label{section3d}
We are after solutions dual to 3d ${\cal N}=4$ super-conformal field theories. Matching the global symmetries of the field theory
implies that the background must have  isometries $\text{SO}(2,3)\times \text{SU}(2)_C\times \text{SU}(2)_H$ and preserve eight Poincar\'e supercharges.
Our geometries must contain an AdS$_4$ factor and a couple of two spheres  $S^2_1(\theta_1,\varphi_1)$ and  $S^2_2(\theta_2,\varphi_2)$. There are two extra directions labelled by $(\sigma,\eta)$.  The  presence of $\text{SO}(2,3)\times \text{SU}(2)_C\times \text{SU}(2)_H$  isometries allow for warp factors that depend only on $(\sigma,\eta)$. The Ramond and Neveu-Schwarz fields must also respect the above-mentioned isometries.
 
The preservation of eight Poincar\'e supersymmetries implies that the generic type IIB background
can be written in terms of a function $V_3(\sigma,\eta)$. 
In string frame the solution reads \cite{Akhond:2021ffz},
\begin{eqnarray}
& & ds_{10,st}^2= f_1(\sigma,\eta)\Big[ds^2(\text{AdS}_4) + f_2(\sigma,\eta) d s^2 (S^2_1)+ f_3(\sigma,\eta) d s^2 (S^2_2)+ f_4(\sigma,\eta)(d\sigma^2+d\eta^2) \Big], \nonumber\\[2mm]
& &e^{-2\Phi}=f_5(\sigma,\eta), \;\; B_2=f_6(\sigma,\eta) \text{Vol}(S^2_1),\;\;C_2= f_7(\sigma,\eta) \text{Vol}(S^2_2),\;\;\; \tilde{C}_4= f_8(\sigma,\eta) \text{Vol(AdS}_4), \nonumber\\ [2mm]
& & f_1=\frac{\pi}{2}\sqrt{\frac{\sigma^3 \partial^2_{\eta \sigma}V_3}{\partial_{\sigma}(\sigma \partial_{\eta} V_3)}},\;\; f_2= -\frac{\partial_\eta V_3 \partial_{\sigma}(\sigma \partial_{\eta} V_3)}{\sigma \Lambda},\;\;f_3= \frac{\partial_{\sigma}(\sigma \partial_{\eta} V_3)}{\sigma \partial^2_{\eta \sigma} V_3},\;\; f_4= -\frac{\partial_{\sigma}(\sigma \partial_{\eta} V_3)}{\sigma^2 \partial_{\eta}V_3},\nonumber\\[2mm]
& & f_5=-16\frac{\Lambda  \partial_{\eta}V_3}{ \partial^2_{\eta\sigma} V_3} , \;\;\;\;  f_6= \frac{\pi}{2} \left(\eta -\frac{\sigma  \partial_{\eta}V_3 \partial_{\eta}^2 V_3}{\Lambda }\right) ,\;\;\;\;  f_7 = -2 \pi \left(\partial_\sigma (\sigma  V_3)-\frac{\sigma  \partial_{\eta}V_3 \partial_{\eta}^2 V_3}{\partial^2_{\eta\sigma} V_3}\right) , \nonumber\\[2mm]
& & f_8 = -\pi^2 \sigma ^2 \left(3 \partial_{\sigma} V_3+\frac{\sigma  \partial_{\eta} V_3 \partial_{\eta}^2 V_3}{\partial_\sigma (\sigma  \partial_{\eta}V_3) }\right), \;\;\;\; \Lambda = \partial_{\eta}V_3\partial^2_{\eta\sigma} V_3 + \sigma \left(( \partial^2_{\eta\sigma} V_3 )^2 + ( \partial^2_{\eta} V_3 )^2 \right). \label{background}
\end{eqnarray}
Where the fluxes are defined from the potentials as follows,
\begin{equation}
F_1=0 , \quad H_3 = d B_2 \quad F_3 = d C_2, \quad F_5= d \tilde{C}_4 + *d \tilde{C}_4 .
\end{equation}
The configuration in eq.(\ref{background}) is solution to the Type IIB equations of motion, if the function $V(\sigma,\eta)$ satisfies,
\begin{equation}
\partial_\sigma \left(\sigma^2 \partial_\sigma V_3\right) +\sigma^2 \partial^2_\eta V_3=0.\label{diffeq} 
\end{equation}
As proven in 
\cite{Akhond:2021ffz} and in detail in Appendix \ref{appendixmapping},
this infinite family of solutions is equivalent to the backgrounds described by D'Hoker, Estes and Gutperle in  \cite{DHoker:2007hhe}. 
\subsubsection{Resolution of the PDE and quantisation of charges}
Following \cite{Akhond:2021ffz}, define $V_3(\sigma,\eta)= \frac{\hat{V}_3(\sigma,\eta)}{\sigma}$ and $\hat{V}_3(\sigma,\eta) = \partial_\eta \hat{W} (\sigma,\eta) $. Consider the coordinates to range in
$0\leq \eta\leq P$, where $P$ is a real number, and $-\infty <\sigma <\infty$.
The differential equation (\ref{diffeq}) must be supplemented by boundary  and initial conditions. In terms of $\hat{W}(\sigma,\eta)$ the problem reads
\begin{eqnarray}
& & \partial^2_\sigma \hat{W}(\sigma,\eta)+\partial^2_\eta \hat{W}(\sigma,\eta)=0, \qquad \qquad \text{(almost everywhere)} \label{PDEhatv}\\
& &   \hat{W} (\sigma,\eta=0)=0,\;\;\;\;\;\hat{W} (\sigma,\eta=P)=0,\nonumber\\
& & \partial_\sigma \hat{W}(\sigma=0^+,\eta)- \partial_\sigma \hat{W}(\sigma=0^-,\eta)=- {\cal R}(\eta).\nonumber
\end{eqnarray}
In analogy with the five dimensional case discussed above, the function ${\cal R}(\eta)$ is the input  determined by the dual quiver field theory. Notice that, since $\hat{W}$ is a harmonic function, we have that also $\hat{V}$ is harmonic, which in turn implies \eqref{diffeq}.

Using a Fourier decomposition for the rank function ${\cal R}(\eta)$ as in the five-dimensional case- see eq.(\ref{rankfunction}), the solution to the problem in eq.(\ref{PDEhatv}) is,
\begin{eqnarray}
& & \hat{V}_3(\sigma,\eta)= \sum_{k=1}^\infty b_k \cos\left( \frac{k\pi\eta}{P}\right) e^{-\frac{k \pi |\sigma|}{P}}, \nonumber\\
& & \hat{W}(\sigma,\eta)= \sum_{k=1}^\infty b_k \left(\frac{P}{k\pi}\right) \sin\left( \frac{k\pi\eta}{P}\right) e^{-\frac{k \pi |\sigma|}{P}}.\label{solutionPDE}\\
& & b_k=\frac{1}{P}\int_0^P {\cal R}(\eta) \sin\left( \frac{k\pi\eta}{P}\right)= \frac{\mathcal{R}_k}{2}.\nonumber
\end{eqnarray}
The study of the quantised charges for Neveu-Schwarz five branes, imposes that the size of the interval $P$ is an integer--consistently with the boundary conditions in eq.(\ref{PDEhatv}), exactly as it occurs in the five dimensional system. Also, in analogy with the 5d-case,  to have quantised numbers of D3 and D5 branes, the rank function must be a piece-wise linear and continuous function of the exact same form as in the five dimensional case--see eq.(\ref{rankfull}).
%
  %
%
%

In the case without offsets, $N_0=N_P=0$, the number of D3   (colour) branes and D5 (flavour) branes in the interval $[k,k+1]$ and the total number of branes are given in \cite{Akhond:2021ffz} ,
\begin{eqnarray}
& & N_{D3}[k, k+1]= N_k,\;\;\;\; N_{D5}[k,k+1]=2 N_k - N_{k+1}- N_{k-1},\label{chargesk}\\
& & N_{D3}^\text{total}= \int_0^P {\cal R}(\eta) d\eta,\;\;\;\; N_{D5}^\text{total}= {\cal R}'(0) - {\cal R}'(P), \;\;\;\; N_{NS5}^\text{total}=P.\nonumber
\end{eqnarray}
The rank function ${\cal R}(\eta)$ encodes the `kinematic data' of the dual conformal field theory. The presence of $P$ NS5 branes along the $\eta$-direction suggests that we should place one NS5 at each integer value of $\eta$. In between the $k^{th}$ and $(k+1)^{th}$ NS5-branes, we have $N_k$ D3 branes and $F_{k}=2 N_k - N_{k+1}- N_{k-1} $ D5 branes as indicated in eq.(\ref{chargesk}). 
%

The ${\cal N}=4$ quiver field theory, for the rank function without offsets is the same as that in the five dimensional system, drawn below eq.(\ref{chargesfinal}). This balanced QFT is proposed to reach a strongly coupled IR fixed point described by the background in eq.(\ref{background}).
\subsection{Summary}
Consider the balanced quiver field theory, preserving eight Poincare SUSYs depicted in the  Figure \ref{fig:quiverfiga}. 
In the case the field theory is  five-dimensional, it is conjectured to reach a strongly coupled fixed point at high energies (compared with the scale defined by the gauge coupling). Conversely, if defined in three dimensions the interacting fixed point will be at low energies.

The quiver can be associated with a rank function. In this section, we consider for generality the situation with offsets as indicated in eq.(\ref{rankfull}),
  %
%
%

We calculate the Fourier coefficient of this rank function using eq.(\ref{rankfunction}). We find,
\begin{eqnarray}
& & \mathcal{R}_k= \frac{2}{k \pi}(N_0 + (-1)^{k+1} N_P)\left[ 1- \frac{P}{k\pi} \sin\left( \frac{k\pi}{P}\right) \right] +\frac{2 P}{k^2\pi^2}\sum_{j=1}^{P-1} F_j \sin\left( \frac{k \pi j}{P}\right).\label{rankgeneral}\\
& & F_1= 2N_1-N_2,\;\;\; F_{P-1}= 2N_{P-1}- N_{P-2},\;\;\;\; F_j= 2 N_j- N_{j+1}- N_{j-1}.\nonumber
\end{eqnarray}

We use the expressions in eqs. (\ref{eq:fourier_vhat}) and (\ref{solutionPDE}) to calculate the potentials $\hat{V}_5(\sigma,\eta)$ and $\hat{V}_3(\sigma,\eta)$ in terms of which the supergravity backgrounds in eqs.(\ref{backgroundrescaled}),(\ref{background})
 are defined. 
 Notice that the input, namely the quiver field theory is the same in both cases ($d=5$ or $d=3$). The differences in the dynamics are encoded in the coefficients $a_k$ and  $b_k$, the potentials $\hat{V}_5$ and $\hat{V}_3$ and the different expressions for the functions $f_i(\sigma,\eta)$ in each of the backgrounds  in eqs.(\ref{backgroundrescaled}), (\ref{background}) respectively.
 
It is convenient to define the complex variable
\begin{equation}
\xi= ~e^{-\frac{\pi}{P} \left[ |\sigma|- i \eta \right]} ,\;\;\; -\frac{P}{\pi}\log|\xi |= |\sigma|,\;\;  e^{-\frac{\pi}{P} \left[ |\sigma|- i (\eta \pm J)   \right]} = \xi e^{\pm\frac{i \pi J}{P}}.\label{xidef}
\end{equation}
 
In terms of this complex variable, we find for the five dimensional $\hat{V}_5(\sigma, \eta)$,
\begin{eqnarray}
& & \hat{V}_5(\sigma,\eta)= \frac{N_0}{\pi^2} \Big[ P ~\text{Im} (~\text{Li}_2( \xi)~)  + \frac{P^2}{2\pi} \text{Re} (~\text{Li}_3( \xi e^{\frac{i \pi}{P} } ) -\text{Li}_3(\xi  e^{-\frac{i\pi}{P}  }   )~)    \Big]\nonumber\\
& &  -\frac{N_P}{\pi^2}\Big[     P~ \text{Im} (~\text{Li}_2(- \xi)~) + \frac{ P^2}{2\pi} ~\text{Re}(~ \text{Li}_3(- \xi  e^{\frac{i\pi}{P} })- \text{Li}_3(-\xi  e^{-\frac{i\pi}{P}   })~)                 \Big] +\nonumber\\
& & \frac{ P^2}{2\pi^3} \sum_{J=1}^{P-1} F_J \text{Re}(~ \text{Li}_3( \xi e^{-\frac{i\pi J}{P} }) - \text{Li}_3(\xi e^{\frac{i \pi J}{P}   }) ~). \label{potentialhat5}
 \end{eqnarray}
 In the case $N_0=N_P=0$ (no offsets) this should be compared with the particular expressions obtained in  \cite{Legramandi:2021uds}.
 
By comparing eqs.(\ref{eq:fourier_vhat}) and (\ref{solutionPDE}) we find that  the three-dimensional quantity $\hat{W}(\sigma,\eta)$ is equal to the five dimensional potential $ \hat{V}_5(\sigma,\eta)$ in eq.(\ref{potentialhat5}). The three dimensional potential $\hat{V}_3(\sigma,\eta)$ is,
\begin{eqnarray}
& &\hat{V}_3(\sigma,\eta)=
 \frac{N_0}{4\pi^2}\Big[ - 2\pi \log[(1-\xi ) (1-  \bar{\xi})] - 2 P~ \text{Im} (~\text{Li}_2( \xi e^{\frac{i \pi}{P} }) - \text{Li}_2( \xi e^{-\frac{i \pi}{P}  }) ~)  \Big]\nonumber \\
& & - \frac{N_P}{4\pi^2} \Big[ - 2\pi \log[ (1+\xi) (1+\bar{\xi}  ) ]   - 2 P~ \text{Im}(~\text{Li}_2(- \xi e^{\frac{i \pi}{P}   }) - \text{Li}_2(-\xi  e^{-\frac{i\pi}{P}  } ) ~) \Big] \nonumber\\
& & + \frac{P}{2\pi^2}\sum_{J=1}^{P-1} F_J \text{Im}(~\text{Li}_2( \xi e^{\frac{ i \pi J}{P} }) - \text{Li}_2( \xi e^{-\frac{i \pi J}{P} } ) ~). \label{potentialhat3} 
\end{eqnarray} 
 The analogy  observed in this section between the problems in five and three dimensions is not a coincidence. An analysis for the free energy, reducing the problem to matrix models,
 was performed by Uhlemann in the five dimensional case \cite{Uhlemann:2019ypp} and by Coccia and Uhlemann in the three-dimensional one \cite{Coccia:2020wtk}. This analysis also reveals the analogies between both problems. In Appendix \ref{appendixmatrix}, we briefly summarise these matrix models and link them to the electrostatic problems we discussed in this section.
\\
In the coming section, we briefly write the formulation of the Wilson Loops in generic antisymmetric representations in the electrostatic language discussed above.

\section{Wilson loops in $d=5$ and $d=3$}\label{wilson5-3}
In this section we re-state, in our electrostatic language, the result obtained in \cite{Uhlemann:2020bek},  \cite{Coccia:2021lpp} for the Wilson loops in a given antisymmetric representation.
The precise mapping used in this translation is given in Appendix \ref{appendixmapping}. After this, we discuss two examples in great detail. This makes interesting connections between different results in the bibliography.

Consider first five dimensional linear quiver gauge theories flowing to a SCFT in the UV. The field theory is realised in the low energy regime of stacks of D5 branes (on which the gauge groups are realised) extending between NS five branes with $(p,q)$ five branes. There are also stacks of D7 branes realising the flavour symmetries of the quiver. 

The Wilson loop is calculated using a D3 probe extending in the time direction and a direction perpendicular to the stack of colour D5 branes. This D3  probe preserves the $SU(2)$ R-symmetry of the SCFT. On this probe, charges of D1 brane and of fundamental string F1 are induced. The F1 extends between the D3 probe and $k$ of the $N_l$ D5 colour branes. The s-rule (Pauli principle) indicates that either one or no F1 stretch between the probe D3 and a give D5. The low energy description of such F1, as found in \cite{Assel:2018rcw}, is  given by a one-dimensional ${\cal N}=4$ conformal quantum mechanics in terms of a massive Fermi multiplet. Integrating out these Fermi multiples as in \cite{Gomis:2006sb} generates the Wilson loop for the group $SU(N_l)$ in the $k$-antisymmetric representation. The charge of D1 brane induced on the D3 indicates the number of NS-five branes the D3 probe has 'gone through', hence is in correspondence with the position of the $SU(N_l$)-colour group.

The holographic calculation of these Wilson loops in the $k$-antisymmetric representation requires the calculation of the on-shell action of a D3 probe that extends along AdS$_2$ inside AdS$_6$ in the background of eq.(\ref{backgroundrescaled}). The two sphere $S^2(\theta,\varphi)$ representing the R-symmetry of the SCFT is also wrapped, hence preserved. An $SO(4)$ isometry  realised in the remaining direction inside AdS$_6$ is also preserved.  There are fluxes switched on the probe D3,
\begin{equation}
{\cal F}_2= f_{el} \text{vol}\text{AdS}_2 + f_{mag} \text{vol}\text{S}^2.\label{fluxes}
\end{equation}
This flux induces the above mentioned charge of F1 and D1 on the D3 probe world-volume. The D1 charge is in correspondence with the position of the $l$-stack of D5 branes, hence the position in the $\eta$-coordinate is $\eta_*=l$. 
The F1 charge is associated with the number $\mathbb{k}$, labelling the representation. This can be thought of as the position $\sigma_*$ in the $\sigma$-direction.

This probe was studied by Uhlemann in \cite{Uhlemann:2020bek}, finding the conditions to preserve SUSY that are imposed on the charges in eq.(\ref{fluxes}). The on-shell action for the probe D3 was found (after a Legendre transform). 

The result for the VEV of Wilson loops for a given $\mathbb{k}$-antisymmetric representation $\wedge$  in the electrostatic language is succinct,
\begin{eqnarray}
& &     \ln\langle W_\wedge\rangle =
    3\pi 
    \sum_{k=1}^\infty \frac{\mathcal{R}_k}{2}\left(\frac{P}{k\pi} \right) \sin\left(\frac{k\pi}{P}\eta^* \right) {e^{-\frac{k\pi}{P}|\sigma^*|}}(\frac{k\pi}{P}|\sigma^*|+1).\label{wilson5d}\\
   & &  N_{\rm D1}= \eta^*, \quad N_{\rm F1}=\sum_{k=1}^\infty \frac{\mathcal{R}_k}{2} \sin\left(\frac{k\pi}{P}\eta^* \right) {e^{-\frac{k\pi}{P}|\sigma^*|}}\operatorname{Sgn}(\sigma^*).\nonumber
\end{eqnarray}
The gauge node for which the Wilson loop is computed is labelled by the position along the quiver $\eta^*=1,2,3,4....$. The antisymmetric representation of the Wilson loop $\mathbb{k}$ coincides with the number of F1 strings. One should determine $|\sigma^*|$ by solving the equation for $N_{F1}=\mathbb{k}$ for the particular value of $\eta^*$ given by the chosen gauge node. These values used in eq.(\ref{wilson5d}) give the VEV of the Wilson loop.

Let us now consider the same type of Wilson loop in the case of three dimensional SCFTs. These 3d low energy SCFTs are realised on stacks of D3 branes that extend between NS five branes. There are also stacks of  D5 branes, representing the flavour groups.  As found in \cite{Assel:2015oxa}, the Wilson loop in the $\mathbb{k}$-antisymmetric representation is calculated by a probe D5' brane, extending along time, preserving  one of the $SU(2)$ of the R-symmetry and also preserving a $U(1)$ part of the R-symmetry. Hence, this probe does not extend  along the same directions of flavour D5 branes.  The probe preserves $SO(1,1)\times SU(2)_R\times U(1)_L$ inside the (bosonic part of) the original symmetry group of the 3d SCFT $SO(2,3)\times SU(2)_L\times SU(2)_R$. It should also preserve four SUSYs.

This probe D5' is introduced in between the NS five branes  that limit the $l$-stack of colour D3 branes. Like in the five dimensional situation above analysed, the parameter $\mathbb{k}$ labelling the antisymmetric representation is realised by the charge of F1 induced on the D5'. A very similar procedure---the integration of a one dimensional Fermi multiplet describing the dynamics of these F1 strings leads to the insertion of a Wilson operator in the $\mathbb{k}$-antisymmetric representation. See 
\cite{Gomis:2006sb}, \cite{Assel:2015oxa}. 

In the holographic dual background of eq.(\ref{background}) the VEV of the Wilson loop is calculated by the on-shell action of a D5' that wraps AdS$_2$--to realise $SO(1,1)$, the two sphere $S^2(\theta_1,\varphi_1)$--to realise the $SU(2)_R$ and a circle inside $S^2(\theta_2,\varphi_2)$. This is achieved by the D5' extending along a direction parametrised by $y$ and choosing $\eta(y)$, $\sigma(y)$, $\theta_2(y)$. Like in the five dimensional case a world-volume flux on the D5' is needed to preserve SUSY, see eq.(\ref{fluxes}).
Imposing SUSY preservation on this probe, Coccia and Uhlemann \cite{Coccia:2021lpp} found the relation between the position of the D5' probe in the $(\sigma,\eta)$ plane and the induced charges of F1 and D3. From there the authors of \cite{Coccia:2021lpp} calculated the Legendre transformed on-shell action for the D5'. 

In our electrostatic language,  we find calculating with the background in eq.(\ref{background}), 
\begin{eqnarray}
& &    \ln\langle W_\wedge\rangle =
     \pi\sum_{k=1}^\infty \frac{{\cal R}_k }{2} \left(\frac{P}{k\pi} \right) \sin\left(\frac{k\pi}{P}\eta^* \right) {e^{-\frac{k\pi}{P}|\sigma^*|}}(\frac{k\pi}{P}|\sigma^*|+{1}),\label{wilson3d}\\
   & &  N_{D3}=\eta^*, \quad N_{\mathrm{F} 1}
   =
    \sum_{k=1}^\infty \frac{{\cal R}_k}{2} \sin\left(\frac{k\pi}{P}\eta^* \right) {e^{-\frac{k\pi}{P}|\sigma^*|}}\operatorname{Sgn}(\sigma^*).\nonumber
\end{eqnarray}
A similar explanation as in the five dimensional case applies here: the value of $N_{D3}=\eta^*=1,2,3,4,....$ labels the gauge node along the quiver. The number of fundamental strings, identified with the $\mathbb{k}$ labelling the antisymmetric representation determine the value of $|\sigma^*|$. These two values used in eq.(\ref{wilson3d}) give the VEV of the Wilson loop. Below, we discuss examples.
  
Notice that both expressions (\ref{wilson5d})-(\ref{wilson3d}) are virtually identical. This is confirmed by the matrix model treatment of these Wilson loops, that shows as discussed in Appendix \ref{appendixmatrix}, that from a field theory viewpoint both expression differ only in proportionality factors. It would be interesting to learn about sub-leading corrections to this result.
Notice also that the result can be written both in five  and in three dimensions, using the fact that the three dimensional quantity $\hat{W}(\sigma,\eta)$  is identical to $\hat{V}_5(\sigma,\eta)$ in five dimensions,
\begin{eqnarray}
& & \mu^{-1 } \ln\langle W_\wedge\rangle =     \hat{V}_5 (\sigma^*,\eta^*)  +{|\sigma^* |}   \operatorname{Sgn} (\sigma^*) N_{F1}    . \label{wilson3d5d}\\
& & \mu_{3d}= \pi,\;\;\;\; ~~~\mu_{5d}= 3\pi.\nonumber
\end{eqnarray}
The expression for $\hat{V}_5 (\sigma^*,\eta^*)$ can be read from eq.(\ref{potentialhat5}), whilst $N_{F1} \operatorname{Sgn}(\sigma_*)$ can be computed using eq.(\ref{rankgeneral}) to be,
\begin{eqnarray}
& & N_{F1} \operatorname{Sgn}(\sigma^*)= -\frac{P}{2\pi^2}   \sum_{J=1}^{P-1} F_J \text{Re}\left[ \text{Li}_2 ( ~ \xi_* e^{\frac{i \pi J}{P} } ~  ) -\text{Li}_2 ( ~\xi _* e^{-\frac{i \pi J}{P}  } ~   )          \right]      \label{eqNF1}\\
& &  + \frac{N_0}{2\pi}\Big[ 2\text{Im} [~\text{Li}_1 ( \xi_* )~] + \frac{P}{\pi} \text{Re}[~ \text{Li}_2 (~\xi_* e^{\frac{i \pi}{P}} ~)-  \text{Li}_2 (~ \xi_*  e^{-\frac{i \pi}{P}  }  ~)~    ] \Big]
\nonumber\\
& &
 -
 \frac{N_P}{2\pi}\Big[ 2\text{Im} [~\text{Li}_1 (  -\xi_*)~] + \frac{P}{\pi} \text{Re}[~ \text{Li}_2 ( -\xi_* e^{\frac{ i \pi}{P}  } ~) -  \text{Li}_2 (-\xi_*  e^{-\frac{i \pi}{P}  }  ~)~    ] \Big]\nonumber
\end{eqnarray}
We have defined $\xi_*= e^{-\frac{\pi}{P} [|\sigma^* | - i\eta^*]}$.

For the reader's convenience we write explicitly eq.(\ref{wilson3d5d}),
\begin{eqnarray}
& & \mu^{-1}  \ln\langle W_\wedge\rangle = \label{wilsonexplicit}\\
& & \frac{P^2}{2\pi^3} \sum_{J=1}^{P-1} F_j \text{Re}\Big[ \text{Li}_3 (  \xi_* e^{-\frac{i \pi J}{P} }   ) -\text{Li}_3 ( \xi _* e^{\frac{i \pi J}{P}  }   ) + \log|\xi_*| \left( \text{Li}_2 ( \xi_* e^{\frac{i \pi J}{P} }   ) -\text{Li}_2 ( \xi _* e^{-\frac{i \pi J}{P}  }   )  \right) \Big]\nonumber\\
& & +\frac{P N_P}{\pi^2} \Big[  - \text{Im}[ ~~ \text{Li}_2 ( - \xi_*  ) - \log|\xi_*|  \text{Li}_1 ( -\xi _*  )~~] +\frac{P}{2\pi}\text{Re} \big[~~
\text{Li}_3 ( - \xi_* e^{-\frac{i \pi }{P} }  ) -\text{Li}_3 ( -\xi _* e^{\frac{i \pi }{P}  }    ) +\nonumber\\
& & +\log|\xi_*| ( ~~\text{Li}_2 (  \xi_* e^{\frac{i \pi }{P} }   ) -\text{Li}_2 ( -\xi _* e^{-\frac{i \pi }{P}  }   )     ~~ ) 
   ~\big]            \Big]+ \nonumber\\
  & & +\frac{P N_0}{\pi^2} \Big[   \text{Im}[ ~~ \text{Li}_2 (  \xi_*  ) - \log|\xi_*|  \text{Li}_1 ( \xi _*  )~~] +\frac{P}{2\pi}\text{Re} \big[~~
\text{Li}_3 (  \xi_* e^{\frac{i \pi }{P} }  ) -\text{Li}_3 ( \xi _* e^{-\frac{i \pi }{P}  }    ) +\nonumber\\
& & -\log|\xi_*| ( ~~\text{Li}_2 (  \xi_* e^{\frac{i \pi }{P} }   ) -\text{Li}_2 ( \xi _* e^{-\frac{i \pi }{P}  }   )     ~~ ) 
   ~\big]            \Big]. \nonumber 
\end{eqnarray}
The reader should compare this expression (in the case of no offsets $N_0=N_P=0$), with the field theoretical expression (obtained with matrix models calculation) in equation (4.53) of \cite{Coccia:2021lpp}. 
\\
In the rest of this section we evaluate in two examples, the expressions for the potentials $\hat{V}_5=\hat{W}$, $\hat{V}_3$ and $ \ln\langle W_\wedge\rangle$ in eqs.(\ref{potentialhat5}),(\ref{potentialhat3}) and (\ref{wilsonexplicit}) respectively.
We focus on the cases of the $T_{M,N}$ and $+_{M,N}$ both in 3d and in 5d. Whilst these are non-generic examples, they are very used in the existing bibliography. We will work them out using rank functions with and without offsets, finding relations between these cases that clarify previous results in the bibliography.

\subsection{Example 1}\label{example1}

Let us consider a five dimensional gauge theory called $\tilde{T}_{N,P}$. The gauge theory is described (in the IR) by the quiver 
\begin{center}
	\begin{tikzpicture}
	\node (1) at (-6,0) [circle,draw,thick,minimum size=1.4cm] {N};
	\node (2) at (-4,0) [circle,draw,thick,minimum size=1.4cm] {2N};
	\node (3) at (-2,0) [circle,draw,thick,minimum size=1.4cm] {3N};	
	\node (4) at (0,0)  {$\dots$};
	\node (6) at (4,0) [rectangle,draw,thick,minimum size=1.2cm] {PN};
	\node (5) at (2,0)  {(P-1)N};
	\draw[thick] (1) -- (2) -- (3) -- (4) -- (5)-- (6);
	\draw[thick] (2,0) circle (0.7cm) ;
	\draw[thick] (1,0) -- (1.3,0);
	\draw[thick] (2.7,0) -- (3.3,0);
	\end{tikzpicture}\
\end{center}
In the case of this same quiver being considered in three dimensions, the above quiver is the UV description of the QFT, flowing to a SCFT in the IR. Let us study all the quantities discussed above.

First, we consider the case {\it without offsets}, the rank function associated  with this quiver is,
\[ {\cal R}(\eta) = \begin{cases} 
N\eta & 0\leq \eta \leq (P-1) \\
N(P-1) (P-\eta)& (P-1)\leq \eta\leq P .
\end{cases}
\]
If the quiver is five dimensional, the number of D7-branes can be read either from $\mathcal{R}'' = NP \delta(\eta-P+1)$, or from eq.\eqref{chargesfinal} which gives $Q_{D7} = P N$.  The number of D5 branes  at the positions
$\eta=1,2,3,4,\text{etc}$,  is the value of $\mathcal{R}(\eta)$ at those points.
This coincides with the ranks of the
first, second, third, fourth node, etc. In total, we have $\int_0^P {\cal R} d\eta=\frac{NP(P-1)}{2}$ D5 branes.  We also have a total of $P$ NS-five branes.

In the three dimensional case, we use eq.(\ref{chargesk}). The number of D5 branes (flavours) is $NP$. The number of D3 branes in each interval coincides with the values of the rank function and there are a total of
$\frac{NP(P-1)}{2}$ D3 branes. The number of NS five branes is also $P$.

Given the rank function above, the  coefficient $\mathcal{R}_k$ can be read from eqs.(\ref{rankfunction}),(\ref{rankgeneral}), with $N_0=N_P=0$ (no offsets) and $F_j= N P \delta_{j, P-1}$. We find,
\begin{equation}
\mathcal{R}_k= \frac{2N P^2 }{k^2\pi^2} \sin\left(\frac{k\pi}{P} \right) (-1)^{k+1}.
\end{equation}
The five dimensional Fourier coefficient $a_k$ as defined in eq.(\ref{eq:fourier_vhat}) and the Fourier coefficient of the three dimensional potential $b_k$ in eq.(\ref{solutionPDE}) are,
\begin{equation}
a_k= (-1)^{k+1}\frac{N P^3}{k^3 \pi^3} \sin\left( \frac{k\pi }{P}\right),\;\;\;\;\;
b_k= \frac{N P^2 }{k^2\pi^2} \sin\left(\frac{k\pi}{P} \right) (-1)^{k+1}. \label{ak1}
\end{equation}

Using eqs.(\ref{eq:fourier_vhat})  and (\ref{potentialhat5}) 
the potential $\hat{V}_5(\sigma,\eta)$ is
\begin{equation}
\hat{V}_5(\sigma,\eta) = \frac{N P^3}{2 \pi ^3} \text{Re} \left(\text{Li}_3(-\xi e^{\frac{ i \pi}{P}  } )-\text{Li}_3(-\xi e^{-\frac{ i \pi}{P} }) \right) \, .\label{eq:V_TN}
\end{equation}
For the three dimensional case, we find $\hat{W}=\hat{V}_5(\sigma,\eta)$ and  using eqs.(\ref{solutionPDE}) and (\ref{potentialhat3})
\begin{equation}
\hat{V}_3(\sigma,\eta)= \frac{N P^2}{2\pi^2}
\text{Im} \left(\text{Li}_2(-\xi e^{-\frac{ i \pi}{P}  })-\text{Li}_2(-\xi e^{\frac{i \pi}{P}  }) \right) \, .\label{V3dTnp}
\end{equation}
We now calculate the Wilson loop using eq.(\ref{wilson3d5d}). The part corresponding to $\hat{V}_5(\sigma^*,\eta^*)$ is read from eq.(\ref{eq:V_TN}). The part corresponding to $N_{F1}$ is calculated explicitly from eqs.(\ref{wilson5d})-(\ref{wilson3d}) or read from eq.(\ref{eqNF1}). We find,
\begin{equation}
N_{F1}\operatorname{Sgn}(\sigma_*)= \frac{NP^2}{\pi^2}\text{Re}\Big[ \text{Li}_2\left(-\xi_*   e^{\frac{i \pi}{P} }  ~)  \right) -  \text{Li}_2\left(   - \xi_* e^{-\frac{i \pi}{P}  }  ~)   \right)   \Big].\label{nf1tnp}
\end{equation}
We write the full result for the Wilson loop, using eq.(\ref{wilsonexplicit}).  In dimensions three ($\mu=\pi$) and in five ($\mu=3\pi$),
\begin{equation}
\mu^{-1} \ln\langle W_\wedge\rangle =\frac{NP^3}{2 \pi^3} \text{Re}\Big[\text{Li}_3(-\xi_* e^{\frac{i\pi}{P}})  - \text{Li}_3(-\xi_* e^{\frac{-i\pi}{P}})  -\log|\xi_* | \left( ~\text{Li}_2(-\xi_* e^{\frac{i\pi}{P}})- \text{Li}_2(-\x_*i e^{\frac{-i\pi}{P}})   ~\right)    \Big].\label{wilsonTn1}
\end{equation}
\\
\\
It is instructive to repeat the calculation for a rank function {\it with offset}. In this case, we choose,
\[ {\cal R}(\eta) = \begin{cases} 
N\eta & 0\leq \eta\leq P .
\end{cases}
\]
Implying $N_0=0, N_P=N P$.
In this case the number of flavour branes will be found from the derivative ${\cal R}'(\eta)= NP \delta(\eta-P+1)$. In other words $F_j=N P\delta_{j, P-1}$.
The Fourier coefficient of the rank function is computed using eq.(\ref{rankfunction}) or equivalently, setting $N_0=0, N_P= PN$ in eq.(\ref{rankgeneral}). The result  is, 
\begin{equation} \label{eq:Trank}
	 \mathcal{R}_{k}=2\frac{NP}{k\pi}(-1)^{k+1}, \quad a_k= \frac{NP^2}{k^2\pi^2}(-1)^{k+1},\;\;\;\;\;
b_k= \frac{NP}{k\pi}(-1)^{k+1}.
\end{equation}
The potential in five dimensions can be calculated using eq.(\ref{eq:fourier_vhat}) or read from eq.(\ref{potentialhat5}). Using the variable $\xi$ defined in eq.(\ref{xidef}), the result is,
\begin{equation}
	\hat{V}_5(\sigma, \eta)=-\frac{P^2 N}{\pi^{2}} \Im \left[\Li_{2}(-{\xi})\right].\label{potential5dTnoffset}
\end{equation}
For the three dimensional potential $\hat{V}_3$ we calculate using eq.(\ref{solutionPDE}) or read from eq.(\ref{potentialhat3}) the result,
\begin{equation}
\hat{V}_3(\sigma,\eta)=  
- \frac{N P}{2\pi} \log[(1+\xi)(1+\bar{\xi})].\label{potential3doffset}
\end{equation}
For the Wilson loop,  both in 3d and in 5d, we find calculating from eqs. (\ref{wilson5d})-(\ref{wilson3d}), or using the generic  result  in eq.(\ref{wilsonexplicit}), 
\begin{equation}
	\begin{aligned}
\mu^{-1} \ln\langle W_\wedge\rangle = -\frac{NP^2}{\pi^{2}} &\{\ln |\xi|[ \operatorname{Arg}(1-\xi_*  )  ]+\Im\left[\Li_2(-{\xi_* })\right]\}.\label{wilsonTNoffset}
\end{aligned}
\end{equation}
The goal of this carefully developed example is to show that if we perform the limit of $P\to\infty$, keeping $\frac{|\sigma|}{P}$ and $\frac{\eta}{P}$ fixed, the results {\it without offsets}
---see eqs.(\ref{eq:V_TN}),(\ref{V3dTnp}),(\ref{wilsonTn1})  at leading order, reproduce the results computed with the rank function {\it with offsets} in eqs.(\ref{potential5dTnoffset}),(\ref{potential3doffset}) and (\ref{wilsonTNoffset}) respectively.

More precisely, eq.(\ref{eq:V_TN}) would lead to eq.(\ref{potential5dTnoffset}) in the the limit of $P\to\infty$
\begin{equation}
\frac{N P^3}{2 \pi ^3} \text{Re} \left(\text{Li}_3(-e^{-\frac{\pi}{P}  (| \sigma |-i-i \eta  )})-\text{Li}_3(-e^{-\frac{\pi}{P}  (| \sigma |+i-i \eta )}) \right) \longrightarrow -\frac{P^2 N}{\pi^{2}} \Im \left[\Li_{2}(-{\xi})\right],
\end{equation}
eq.(\ref{V3dTnp}) will give eq.(\ref{potential3doffset})
\begin{equation}
\frac{N P^2}{2\pi^2}
\text{Im} \left(\text{Li}_2(-e^{-\frac{\pi}{P}  (| \sigma |+i-i \eta  )})-\text{Li}_2(-e^{-\frac{\pi}{P}  (| \sigma |-i-i \eta )}) \right) \longrightarrow  -  \frac{N P}{2\pi} \log[(1+\xi)(1+\bar{\xi})],
\end{equation}
and eq.(\ref{wilsonTn1}) will result in  eq.(\ref{wilsonTNoffset})
\begin{align}
\frac{NP^3}{2 \pi^3} \text{Re}& \Big[\text{Li}_3(-\xi e^{\frac{i\pi}{P}})  - \text{Li}_3(-\xi e^{\frac{-i\pi}{P}})  -\log|\xi | \left( ~\text{Li}_2(-\xi e^{\frac{i\pi}{P}})- \text{Li}_2(-\xi e^{\frac{-i\pi}{P}})   ~\right)    \Big]\longrightarrow\nonumber\\
 & -\frac{NP^2}{\pi^{2}} \Big[ \ln |\xi|[\operatorname{Arg}(1- {\xi })]+\Im\left[\Li_2(-{\xi})\right]\Big].\label{correspondences}
\end{align}

%
This observation is interesting as it explains differences in results for the potentials $\hat{V}_5$ found in \cite{Uhlemann:2019ypp}  (that used rank functions with offsets) compared with those in \cite{Legramandi:2021uds} (that used rank functions without offsets). This also explains some differences for the Wilson loops VEV computed in \cite{Uhlemann:2020bek} (with offsets) compared with  some of the field theoretical results in \cite{Coccia:2021lpp} (without offsets).
Interestingly, the free energy (or holographic central charge) does not differ at leading order for rank functions with or without offsets; the difference appears only at sub-leading orders in the length of the quiver $P$.
 \subsubsection{A numerical study}
\begin{figure}[h!]
    \centering
    \includegraphics[width=13cm]{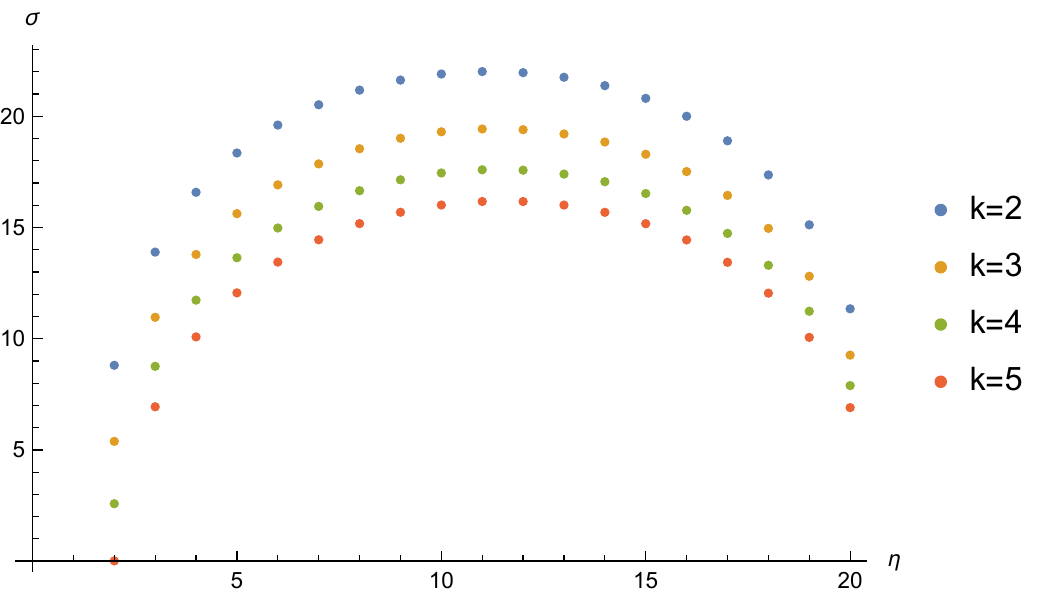}%
\caption{The values of $\sigma^*$ for Wilson loops in the representation $\mathbb{k}=2,3,4,5.$.}
\label{sigmae}
\end{figure}

\begin{figure}[h!]
 \centering
    \includegraphics[width=13cm]{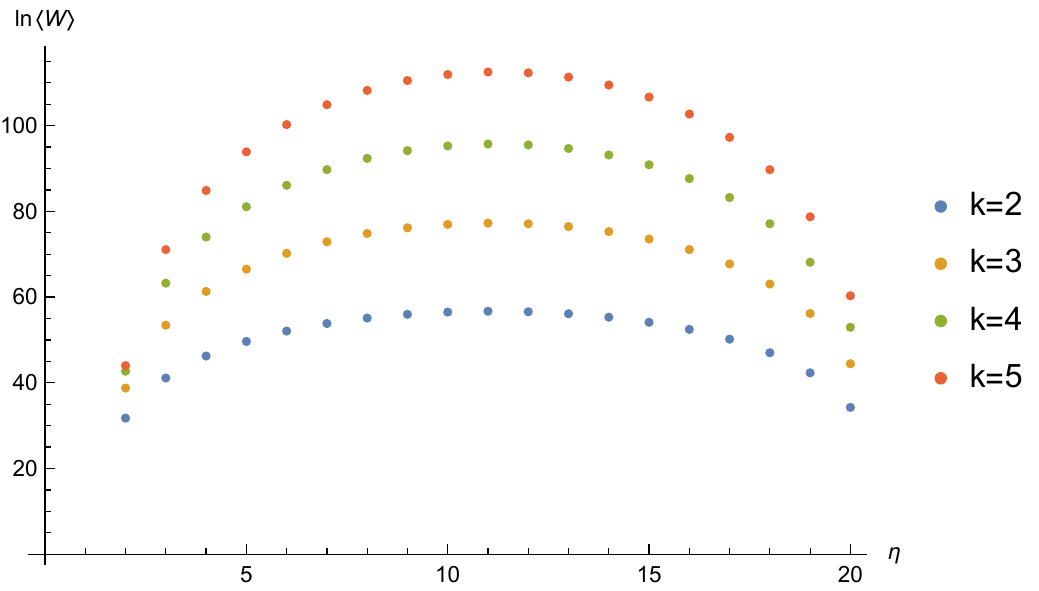} %
\caption{The value of  $ \ln\langle W_\wedge\rangle$. The colour code is the same as in Figure \ref{sigmae}.}
\label{lnWTfig}
\end{figure}

\begin{figure}[h!]
    \centering
    \includegraphics[width=14cm]{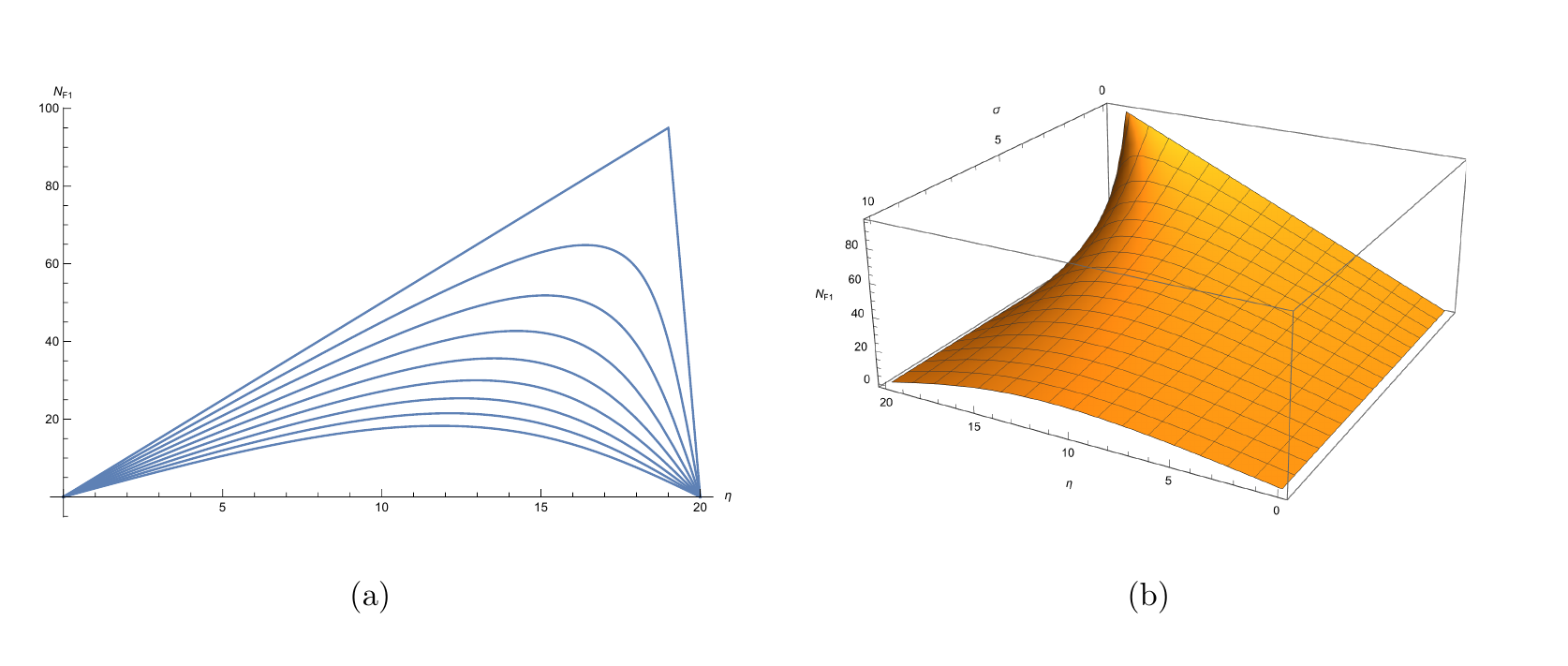} %
\caption{The values of $\sigma^*$ for Wilson loops in the representation $\mathbb{k}=2,3,4,5.$}
\label{WilsonlinePlots}
\end{figure}

 For the benefit of the reader, we discuss an example of the calculation of $\log<W_{\mathbb{k}}>$ in numerical detail.
 Let us consider the $\tilde{T}_{N,P}$ example of this section (with no offsets) for the particular values
 \begin{eqnarray}
 & & N=10,\;\;\; P=20,\nonumber\\
 & & \text{The quiver is}~~ SU(10)\times SU(20)\times....\times SU(190)~~\text{with flavour group}~~ SU(200).\nonumber
 \end{eqnarray}
In the three dimensional case, the groups are unitary $U(N)$. 
 We calculate the Wilson loop for the $l$-gauge group (for $l=1,2,3,4,...,19$) in the $\mathbb{k}$-antisymmetric representation for the values $\mathbb{k}=2,3,4,5$.
 For this, we need to count $N_{F1}=2,3,4,5$ fundamental strings extending between the probe and the stack of colour branes. The position of the probe in the $\eta$-coordinate is $\eta^*=l$. The position in the $\sigma$-coordinate  is obtained by solving eq.(\ref{nf1tnp}) for $\sigma^*$. The numerical solution for different values of $\eta^*=1,...19$ is plotted in Figure \ref{sigmae}. 
Choosing a concrete $\eta^*$ we find the value of $\sigma^*$  making $N_{F1}=2,3,4,5$ colour coded in the figure. The result for the VEV of the Wilson loop is obtained by replacing these values  $(\sigma^*,\eta^*)$ in eq.(\ref{wilsonTn1})--remind that $\xi_*= e^{-\frac{\pi}{P}\left( |\sigma^*|- i \eta^* \right)  } $. The results are shown in Figure \ref{lnWTfig}.
 It is also instructive to plot $N_{F1}(\sigma^*,\eta^*)$ in terms of $\eta^*$ for different values of $\sigma^*$. This is found in inset (a) of Figure \ref{WilsonlinePlots}. In fact, for $\sigma^*=0$ we see the function $\frac{{\cal R}(\eta^*)}{2}$, while for $\sigma^*\to\infty$ we find a vanishing value. This is in agreement with eqs.(\ref{wilson5d}),(\ref{wilson3d}).
 The inset (b) of the same figure shows $N_{F1}(\sigma^*,\eta^*)$. Only the integer values of $N_{F1}$ are acceptable (as this coincides with the representation) for integer values of $\eta^*$ (indicating the gauge node).
%
Let us discuss a different example in a more succinct fashion.

 \subsection{Example 2}\label{example2}
We consider a second  example, known as the $+_{P,N}$ theory. 
We start, as above with discussion {\it without offsets}. The rank function is,
\[ {\cal R}(\eta) = \begin{cases} 
N\eta & 0\leq \eta \leq 1 \\
N & 1\leq \eta\leq (P-1)\\
N (P-\eta) & (P-1)\leq \eta\leq P .
\end{cases}
\]
In the five dimensional case, we have $N$ D7-branes  localised at $\eta = 1$ (the first gauge group) and $N$ D7 branes at $\eta= P-1$ (the last gauge group). This follows from $\mathcal{R}'' = N \delta(\eta-1)+N \delta(\eta-(P-1))$. There are a total of $(P-1)N$ D5-branes, as calculated by $\int_0^P {\cal R} d\eta$. The number comes from $N$ D5 branes for each integer value of $\eta$ between $[1,P-1]$. 
In the three dimensional case, we have $N$ D5-branes  localised at $\eta = 1$ and $N$ D5 branes at $\eta= P-1$, a total of $(P-1)N$ D3-branes and $P$ NS five branes.

This is equivalent to a linear quiver field theory (in 5d it flows to a SCFT in the UV, in 3d it flows to a SCFT in the IR),
\begin{center}
	\begin{tikzpicture}
	\node (1) at (-4,0) [rectangle,draw,thick,minimum size=1.2cm] {N};
	\node (2) at (-2,0) [circle,draw,thick,minimum size=1.4cm] {N};
	\node (3) at (0,0)  {$\dots$};
	\node (5) at (4,0) [rectangle,draw,thick,minimum size=1.2cm] {N};
	\node (4) at (2,0) [circle,draw,thick,minimum size=1.4cm] {N};
	\draw[thick] (1) -- (2) -- (3) -- (4) -- (5);
	\draw [decorate,decoration={brace,amplitude=15pt,mirror},thick,yshift=-1.5em]
	(-2.8,0) -- (2.8,0) node[midway,yshift=-2.5em]{P-1};
	\end{tikzpicture}
\end{center}
We calculate the Fourier coefficients, 
\begin{eqnarray}
& & \mathcal{R}_{k}=2\frac{N P}{k^2 \pi^2} \sin\left( \frac{k\pi }{P}\right) \left(1+(-1)^{k+1}\right),\;\;\;
a_k= \frac{N P^2}{k^3 \pi^3} \sin\left( \frac{k\pi }{P}\right) \left(1+(-1)^{k+1}\right),\nonumber\\
& & b_k= \frac{N P}{k^2 \pi^2} \sin\left( \frac{k\pi }{P}\right) \left(1+(-1)^{k+1}\right).\label{ak2}
\end{eqnarray}
which leads to the potentials
\begin{eqnarray}
& &\hat{V}_5(\sigma,\eta)= \frac{N P^2}{2 \pi ^3} \text{Re}\Big[ \text{Li}_3( \xi e^{-\frac{i\pi}{P} }  ) -\text{Li}_3( \xi e^{\frac{i\pi}{P} } )  + \text{Li}_3(- \xi  e^{\frac{i\pi}{P} } )-\text{Li}_3(-\xi  e^{-\frac{i\pi}{P} })  \Big] \, , \label{eq:+MN}\\
& & \hat{V}_3(\sigma,\eta)= \frac{N P}{2 \pi ^2} \text{Im}\Big[   \text{Li}_2( \xi e^{\frac{i\pi}{P} }  ) - \text{Li}_2( \xi e^{-\frac{i\pi}{P} } )  + \text{Li}_2(- \xi  e^{-\frac{i\pi}{P} } )-\text{Li}_2(-\xi  e^{\frac{i\pi}{P} })  \Big] \, .\nonumber
\end{eqnarray}
Finally, the result for the VEV of the Wilson loop is,
\begin{eqnarray}
& & \mu^{-1} \ln\langle W_\wedge\rangle =\frac{NP^2}{2 \pi^3} \text{Re}\Big[
\text{Li}_3(\xi_* e^{ -\frac{i\pi}{P}})  -
 \text{Li}_3(\xi_* e^{\frac{i\pi}{P}})  +  
 \text{Li}_3(-\xi_* e^{ \frac{i\pi}{P}})  -
 \text{Li}_3(-\xi_* e^{-\frac{i\pi}{P}}) +\nonumber\\
 & & 
 +\log|\xi | \left( ~\text{Li}_2(\xi_* e^{\frac{i\pi}{P}})- \text{Li}_2( \xi_* e^{\frac{-i\pi}{P}})   + ~\text{Li}_2(- \xi_* e^{-\frac{i\pi}{P}})- \text{Li}_2( -\xi_* e^{-\frac{i\pi}{P}}) ~\right)    \Big].\label{wilsonTn}
\end{eqnarray}
\\
We now work out the same results for the situation {\it with offsets}. We have
\[ {\cal R}(\eta) = \begin{cases} 
N & 0\leq \eta\leq P .
\end{cases}
\]
In this case we set $F_j= N\delta_{j,1} + N\delta_{j,P-1}$ and $N_0=N_P=N$.
The Fourier coefficients of the rank function and the potentials in 5d and 3d are,
\begin{eqnarray}
& & 	 \mathcal{R}_{k}= 2\frac{N}{k\pi}\left(1+(-1)^{k+1}\right),\;\; a_k= \frac{P N}{k^2\pi^2} \left(1+(-1)^{k+1}\right),\;\; b_k= \frac{N}{k\pi} \left(1+(-1)^{k+1}\right).
\end{eqnarray}
The potentials read
\begin{eqnarray}
& & 	\hat{V}_5(\sigma, \eta)=\frac{P N}{\pi^{2}} \Im \left[ -\Li_{2}(-\xi)+\Li_{2}( \xi)\right] ,\;\;\;\; \hat{V}_3(\sigma,\eta)= \frac{N}{2\pi}\log\left[ \frac{(1+\xi)(1-\bar{\xi})}{(1-\xi)(1+\bar{\xi})}\right].\label{V+}
\end{eqnarray}
For the Wilson loop, we find
\begin{eqnarray}
& & 
\mu^{-1} \ln\langle W_\wedge\rangle = \frac{P N}{\pi^{2}} \{\ln |\xi_* | [\operatorname{Arg}(1-\xi_*)-\operatorname{Arg}(1+\bar{\xi}_*)]+\Im\left[\Li_{2}(\xi_*)-\Li_2(-{\xi_*})\right]\}.\label{wilson+off}
\end{eqnarray}
As pointed out in the first example, in the limit $P\to\infty$, keeping the quotients $\frac{|\sigma|}{P}$ and $\frac{\eta}{P}$, we find that the results {\it without offsets} in eqs.(\ref{eq:+MN}),(\ref{wilsonTn}) at leading order in $1/P$ are approximated  by the results {\it with offsets} in eqs.(\ref{V+}) and (\ref{wilson+off}).

\section{Mirror symmetry and Wilson loops}\label{mirrorsection}
The material in this section applies primarily to the 3d SCFTs and their string duals in eq.(\ref{background}). The reader might want to extend these results also to the five dimensional case.

It is known that  three dimensional SUSY QFTs with eight supercharges enjoy a symmetry called Mirror symmetry. The idea is that given two partitions $\rho$ and $\hat{\rho}$ of the number $M$ the two theories $T_\rho^{\hat{\rho}} [SU(M)]$ and $T_{\hat{\rho}}^\rho [SU(M)]$ are conjectured to flow to the same IR SCFT. In the language of Hanany-Witten setups, mirror symmetry is realized as an S-duality (we will discuss more about this in Appendix \ref{appS}).

In contrast, in the electrostatic language  described in Section \ref{section3d}, the manifestation of mirror symmetry is in terms of a diffeomorphism; see the paper \cite{Akhond:2021ffz} for explanations. In fact, as explained in \cite{Akhond:2021ffz} for {\it balanced}
quivers with {\it one flavour node}
the mirror is also balanced and with one flavour node, hence suitable to be described by the language in Section \ref{section3d}. 

In what follows, we analyse the effect of a mirror symmetry transformation on the VEV of a Wilson loop in the antisymmetric $\wedge$-representation We show below that if the {\it electric} theory has $N_f^{el}$ flavours and the {\it magnetic} theory has $N_{f}^{mag}$ flavours (in both cases with a single flavour node) the calculation of the Wilson line in the same representation $\wedge$ satisfies,
\begin{equation}
N_f^{el} \ln\langle W^{el}_\wedge\rangle=N_f^{mag} \ln\langle W^{mag}_\wedge\rangle.\label{wilsonrelation}
\end{equation}
To see this, let us consider a {\it generic, balanced, one flavour node} linear quiver field theory. The quiver,  rank function (with no offsets) and Fourier coefficient of this electric theory are
\footnote{For the generic triangular rank function the quotient $\frac{N }{ (P-S)}$ is taken to be integer.},
\begin{center}
             \begin{tikzpicture}
                      \node[label=below:{$N$}][u](N1){};
                      \node[label=below:{$2N$}][u](N2)[right of=N1]{};
                      \node (dots)[right of=N2]{$\cdots$};
                      \node[label=below:{$SN$}][u](NP-1)[right of=dots]{};
                      \node[label=above:{$\frac{SN(P-S-1)}{P-S}$}][u](P-S-1)[right of=NP-1]{};
                      \node[label=below:{$\frac{SN(P-S-2)}{P-S}$}][u](P-S-2)[right of=P-S-1]{};
                      \node (dots')[right of=P-S-2]{$\cdots$};
                      \node[label=below:{$\frac{SN}{P-S}$}][u](rightmost)[right of=dots']{};
                      \node[label=above:{$\frac{NP}{P-S}$}][uf](FP-1)[above of=NP-1]{};
                      \draw(N1)--(N2);
                      \draw(N2)--(dots);
                      \draw(dots)--(NP-1);
                      \draw(NP-1)--(FP-1);
                      \draw(NP-1)--(P-S-1);
                      \draw(P-S-1)--(P-S-2);
                      \draw(P-S-2)--(dots');
                      \draw(dots')--(rightmost);
             \end{tikzpicture}
\end{center}
\[ {\cal R}_e(\eta) = \begin{cases} 
          N \eta & 0\leq \eta \leq S \\
          \frac{N S}{(P-S)}(P-\eta) & S \leq \eta\leq P ,
       \end{cases}
    \]
and the Fourier coefficient of the rank function,
\begin{equation}
\mathcal{R}^e_k= \frac{2N P^2}{(P-S) \pi^2 k^2}\sin\left( \frac{k\pi S}{P}\right).\label{akex1}
\end{equation}
Using eq.(\ref{wilson3d}), the VEV of the Wilson line for this quiver is 
\begin{equation}
 \ln\langle W^{elec}_\wedge\rangle =\pi 
    \sum_{k=1}^\infty \left(\frac{N P^3}{(P-S) \pi^3 k^3} \right) \sin\left( \frac{k\pi S}{P}\right) \sin\left(\frac{k\pi}{P}\eta^* \right) {e^{-\frac{k\pi}{P}|\sigma^*|}}(\frac{k\pi}{P}|\sigma^*|+1).\label{wilsonelec}
\end{equation}
Finally, the combination $N_f^{el} \ln\langle W^{el}_\wedge\rangle$ is
\begin{equation}
N_f^{el} \ln\langle W^{elec}_\wedge\rangle =\pi 
    \sum_{k=1}^\infty \left(\frac{N^2 P^4}{(P-S)^2 \pi^3 k^3} \right) \sin\left( \frac{k\pi S}{P}\right) \sin\left(\frac{k\pi}{P}\eta^* \right) {e^{-\frac{k\pi}{P}|\sigma^*|}}(\frac{k\pi}{P}|\sigma^*|+1).\label{wilsonelecNfe}
\end{equation}
Now, let us perform the same analysis for the mirror magnetic system.
Following the prescription in \cite{Akhond:2021ffz},  the mirror system is described by a magnetic quiver, rank function and Fourier coefficient,

\begin{center}
\begin{scriptsize}
\begin{tikzpicture}
\node[label=above:{$P-S$}][u](1){};
\node[label=below:{$2(P-S)$}][u](2)[right of=1]{};
\node(dots)[right of=2]{$\cdots$};
\node[label=above:{$S(N+1)-P$}][u](3)[right of=dots]{};
\node[label=below:{$SN$}][u](mid)[right of=3]{};
\node[label=above:{$S(N-1)$}][u](3')[right of=mid]{};
\node(dots')[right of=3']{$\cdots$};
\node[label=below:{$S$}][u](1')[right of=dots']{};
\node[label=above:{$P$}][uf](f)[above of=mid]{};
\draw(1)--(2);
\draw(2)--(dots);
\draw(dots)--(3);
\draw(3)--(mid);
\draw(mid)--(f);
\draw(mid)--(3');
\draw(3')--(dots');
\draw(dots')--(1');
\end{tikzpicture}
\end{scriptsize}	
\end{center}

\begin{equation}
\mathcal{R}_m(\hat{\eta})=\left\{\begin{array}{cc}
(P-S)\hat{\eta} & \hat{\eta}\in[0,\frac{SN}{P-S}]  \\
S\left(\frac{NP}{P-S}-\hat{\eta}\right)& \hat{\eta}\in[\frac{SN}{P-S},\frac{NP}{P-S}]
\end{array}\right.\label{figureXX3}
\end{equation}

\begin{equation}
\mathcal{R}_k^{(m)}= \frac{ 2(P-S) }{ NP}\int_0^{\frac{NP}{P-S} } \mathcal{R}^{m}(\hat{\eta}) \sin \left( \frac{k\pi(P-S) \hat{\eta}}{NP} \right) d\hat{\eta}= \frac{2NP^2 }{ (P-S)\pi^2 k^2 }\sin\left( \frac{k\pi S}{P}\right).\label{akm}
\end{equation}
Notice that
$\mathcal{R}_k^{(e)}=\mathcal{R}_k^{(m)}$. Also, note that the range of the 'electric' coordinate is $0\leq \eta\leq P$, whilst for the 'magnetic' coordinate we have $0\leq \hat{\eta}\leq \frac{NP}{P-S}  $. Finally, observe that in both mirror descriptions 
\begin{eqnarray}
& &\text{electric description:}~~~N_{NS5}=P,\;\;\; N_{D5}=\frac{N P}{(P-S)},\nonumber\\
& & \text{magnetic description:}~~~N_{NS5}=\frac{NP}{(P-S)},\;\;\;\; N_{D5}=P.\nonumber
\end{eqnarray}
Following  \cite{Akhond:2021ffz},  we perform the  identifications between electric variables ($\sigma,\eta$) and their magnetic counterparts $(\hat{\sigma},\hat{\eta})$,
\begin{eqnarray}
& & \eta\leftrightarrow \frac{N_{NS5}}{N_{D5}} \hat{\eta}=\frac{(P-S)}{N} \hat{\eta},~~\sigma \leftrightarrow \frac{N_{NS5}}{N_{D5}} \hat{\sigma}=\frac{(P-S)}{N} \hat{\sigma}.\label{identifmirror}
\end{eqnarray}
The VEV of the magnetic  Wilson loop is found by carefully using eq.(\ref{akm}) in eq.(\ref{wilson3d}),
\begin{equation}
 \ln\langle W^{mag}_\wedge\rangle =\pi 
    \sum_{k=1}^\infty \left(\frac{N^2 P^3}{(P-S)^2 \pi^3 k^3}\right) \sin\left( \frac{k\pi S}{P}\right) \sin\left(\frac{k\pi(P-S)}{PN}\hat\eta^* \right) {e^{-\frac{k\pi(P-S))}{PN}|\hat\sigma^*|}}\left(\frac{k\pi(P-S)}{PN}|\hat\sigma^*|+1 \right).\label{wilsonmag}
\end{equation}
As above, the combination
\begin{align}
N_f^{mag} &\ln\langle W^{mag}_\wedge\rangle= \nonumber\\
&\pi 
    \sum_{k=1}^\infty \left(\frac{N^2 P^4}{(P-S)^2 \pi^3 k^3}\right) \sin\left( \frac{k\pi S}{P}\right) \sin\left(\frac{k\pi(P-S)}{PN}\hat\eta^* \right) {e^{-\frac{k\pi(P-S))}{PN}|\hat\sigma^*|}}\left(\frac{k\pi(P-S)}{PN}|\hat\sigma^*|+1 \right).\label{wilsonmagNfmag}    
\end{align}
For a given electric node (labelled by $\eta^*$) in a given  antisymmetric representation (labelled by $\wedge$),  we find the Wilson loop in eq.(\ref{wilsonelec}). We compare this with the magnetic Wilson loop in the same $\wedge$-representation  calculated for a  {\it different} magnetic node labelled by $\hat{\eta}^*$. If these nodes satisfy
\begin{equation}
\text{electric node}= \frac{P-S}{N}\times \text{magnetic node}\longrightarrow \eta_*=\frac{P-S}{N}\hat{\eta}_*,\nonumber
\end{equation}
then, using the identification in eq.(\ref{identifmirror}), we find that eq.(\ref{wilsonelecNfe}) equals eq.(\ref{wilsonmagNfmag}), which is the relation in eq.(\ref{wilsonrelation}). 

Another way to arrive to eq.(\ref{wilsonrelation}) is to start from eq.(\ref{wilsonexplicit}). Notice that for the electric and magnetic quivers above, we have
\begin{eqnarray}
& & \xi_{*,el}= e^{-\frac{\pi}{P} \left(|\sigma^*_{el}| -i\eta^*_{el}\right)},\;\;\; F_J= \delta_{J,S} \frac{N P}{(P-S)},\;\;\; \;\;\;\;\text{quiver length}=P.\label{relacionesEM}\\
& & \hat{\xi}_{*,mag}=e^{-\frac{\pi (P-S)}{P N} \left(|\hat{\sigma}^*_{mag}| -i\hat{\eta}^*_{mag}\right)}= \xi_{*,el},\;\;\; F_J= \delta_{J,\frac{SN}{(P-S)}} P,\;\;\; \text{quiver length}=\frac{N P}{(P-S)}.\nonumber
\end{eqnarray}
We have used the rescaling in eq.(\ref{identifmirror}). Then, using eq.(\ref{wilsonexplicit})--for the case of no offsets, $N_0=N_P=0$, and multiplying the electric (magnetic) result by the electric (magnetic) number flavours we get eq.(\ref{wilsonrelation}).

Within the class of balanced linear quiver with one flavour node discussed in this section, one special subclass is those quivers that are self-mirror. They are characterised by the condition $N=(P-S)$. This implies that electric result in eq.(\ref{wilsonelec}) and the magnetic result in eq.(\ref{wilsonmag}) are identical (same node, same representation, same number of flavours) and eq.(\ref{wilsonrelation}) is automatically satisfied.

\section{Conclusions}\label{concl}
Let us start with a brief overview of the contents of this paper. In Section \ref{sectiongeometry} we summarised  the electrostatic description of an infinite family of Type IIB backgrounds dual to SCFTs in five and three spacetime dimensions, preserving eight Poincare supercharges. The electrostatic  point of view is complementary to the 'holomorphic' one developed in \cite{DHoker:2007hhe},\cite{DHoker:2016ujz}, that we review carefully in Appendix \ref{appendixmapping}. The electrostatic description of Section \ref{sectiongeometry} makes the connection with quantum field theory (in particular, with the matrix-model/localisation) more transparent. We discuss this briefly in Appendix \ref{appendixmatrix}.

In Section \ref{wilson5-3} we write (using the formalism of Section \ref{sectiongeometry}) the result for the VEV of Wilson loops for a given gauge group in a given antisymmetric $\wedge$-representation. We make clear that this observable takes the same expression in five and in three spacetime dimensions. Whilst this was already clear from a purely field theoretical/matrix model perspective \cite{Coccia:2021lpp}, from the holographic point of view, this becomes more transparent, when written in the electrostatic formalism.
Two examples were worked out in full detail to address a somewhat unclear situation in the existing bibliography. In fact, we showed the relation between results when the rank function is taken with (without) offsets is a limit procedure. This translates in field theory language to the presence of offsets in the matrix model eigenvalue distribution. We  have resolved this potentially unclear issue in the bibliography.

In Section \ref{mirrorsection}, we have discussed the action of mirror symmetry on three dimensional quiver field theories and how the holographic description of balanced quivers with one flavour node realises this symmetry. This leads us to propose a relationship between the Wilson loops; the one computed in a given antisymmetric representation for a certain gauge node in the electric description becomes equivalent (up to a precise multiplicative factor) with the Wilson loop in the same representation for a different gauge node in the magnetic description of the same system.

The  very detailed  Appendix \ref{appendixmapping} is of special note. There, we show many details and derivations of the map between the holomorphic and the electrostatic description of these systems, with worked out examples. This might prove useful for colleagues working on these topics.

For the future, it would be interesting to extend the study of Wilson loops to other systems in different dimensions, also admitting electrostatic description. In fact, for SCFTs in dimension six--see \cite{sixd}, four--see \cite{fourd}, two--see \cite{twod} and one--see \cite{oned} for a small sample of references, there is a well developed electrostatic formalism.
Also, the calculation of Wilson loops in symmetric representations or products of various representations seems a feasible problem to study. It should also be nice to further study the relation between Wilson Loops in both mirror descriptions, expressed by eq.(\ref{wilsonrelation}).

We hope that the 'translation character' of this work can show the analogies between the problem in different dimensions. We anticipate that other analogies, similar to those pointed out in this paper, will be encountered when discussing different observables.  We believe that the formalism of Section \ref{sectiongeometry}, and references \cite{sixd}, \cite{Legramandi:2021uds}, \cite{fourd}, \cite{Akhond:2021ffz},  \cite{twod} and  \cite{oned} is the best suited to look for coincidences in observables for systems in diverse dimensions. The analogies suggested by the holographic studies, in turn, may find a field theoretical understanding of their own.

 \section*{Acknowledgments: }
We would like to thank useful discussions with Andrea Legramandi and Christoph Uhlemann. 
We are supported by  STFC  grant  ST/T000813/1 and ST/V507143/1. The authors have applied to a Creative Commons
Attribution (CC BY) licence.

\appendix
 \section{Summary of the Matrix Model associated with the SCFTs}\label{appendixmatrix}

In this appendix, we discuss  matrix model calculations of certain observables in QFTs that flow to SCFTs. We will phrase quantities in the electrostatic language using harmonic potentials. Five and three dimensional linear quiver QFTs flowing to SCFTs are considered below. This appendix is a (very) brief summary of certain aspects of the paper \cite{Akhond2}.

Various quantities in 5d SCFTs are studied using matrix model calculations in \cite{Uhlemann:2019ypp,Uhlemann:2020bek}. One can check that the free energy calculations of the theory in terms of the function $\varrho(z,x)=N(z)\hat\rho(z,x)$, satisfying the saddlepoint equation 
\begin{equation}
\frac{1}{4} \partial_{x}^{2} \varrho(z, x)+\partial_{z}^{2} \varrho(z, x)+P^{2} k(z) \delta(x)=0,
\end{equation}
can be related to our language with the mapping (explained in \cite{Akhond2}),
\begin{equation}
	N(z)\hat\rho(z,x)=2P\partial^2_\eta \hat{V}_5(\sigma,\eta) \quad z=\eta/P \quad x=\sigma/(2P) \quad N(z)=\cal{R}(\eta).
\end{equation}

For 5d antisymmetric Wilson loops with association with gauge node at location $z$ in the quiver in $\mathbb{k}$-antisymmetric rank, using these relations, the matrix model calculation would read
\begin{align}
	 \ln\langle W_\wedge\rangle &= -6 \pi P N(z) \int_{b(0, z)}^{b(\mathbb{k}, z)} b \hat{\rho}(z, b) d b \equiv -6 \pi P \int_{b(0, \eta)}^{b(\mathbb{k}, \eta)} \sigma/(2P) 2P\partial^2_\eta \hat V(\sigma,\eta) d \sigma/2P \\
	 &=-3 \pi (\int_{b(0, \eta)}^{b(\mathbb{k}, \eta)} -\partial_\sigma \hat V(\sigma,\eta)d \sigma+\left.\sigma \partial_\sigma \hat V(\sigma,\eta)\right|^{(\sigma*,\eta*)}_\infty)=-3 \pi (- \hat {V}_5(\sigma,\eta)+\left.\sigma \partial_\sigma \hat{V}_5(\sigma,\eta))\right|_{(\sigma*,\eta*)}.
\end{align}
The Laplace equation and by part integration are used in the second line. The above equation must be evaluated at the point $ \sigma^*=b(\mathbb{k}, \eta^*)$ which is determined by the integral equation
\begin{equation}
	\mathbb{k} \equiv N(z)y=N(z)\int_{b(z, y)}^{\infty} d x \hat{\rho}(z, x) \equiv  \int_{b(0, \eta)}^{b(\mathbb{k}, \eta)} 2P\partial^2_\eta \hat V(\sigma,\eta) d \sigma/2P = \left.\partial_\sigma \hat{V}_5(\sigma,\eta)\right|^{(\sigma*,\eta*)},
\end{equation}
Hence, by choosing the representation  $\mathbb{k}$ and the gauge node, $(\sigma*,\eta*)$ will be determined, and the Wilson loop can be calculated.

For the 3d field theory calculations \cite{Coccia:2020wtk,Coccia:2021lpp} with a similar map 
\begin{equation}
	N(z)\hat\rho(z,x)=2P\partial^2_\eta \hat V(\sigma,\eta) \quad z=\eta/P \quad x=\sigma/(2P) \quad N(z)=\cal{R}(\eta),
\end{equation}
we have 
\begin{align}
	 \ln\langle W_\wedge\rangle &= 2 \pi P N(z) \int_{b(0, z)}^{b(\mathbb{k}, z)} b \hat{\rho}(z, b) d b \equiv 2 \pi P \int_{b(0, \eta)}^{b(\mathbb{k}, \eta)} \sigma/(2P) 2P\partial^2_\eta \hat W(\sigma,\eta) d \sigma/2P \\
	 &= \pi (- \hat W(\sigma,\eta)+\left.\sigma \partial_\sigma \hat W(\sigma,\eta))\right|^{(\sigma*,\eta*)},
\end{align}
and a similar condition for evaluation point
\begin{equation}
	\mathbb{k} \equiv N(z)y=N(z)\int_{b(z, y)}^{\infty} d x \hat{\rho}(z, x) \equiv  \int_{b(0, \eta)}^{b(\mathbb{k}, \eta)} 2P\partial^2_\eta \hat V(\sigma,\eta) d \sigma/2P = \left.\partial_\sigma \hat W(\sigma,\eta)\right|_{(\sigma*,\eta*)}.
\end{equation}
Again, by choosing the representation  $\mathbb{k}$ and the gauge node, $(\sigma*,\eta*)$ will be determined.

\section{Map between holomorphic and electrostatic formalisms}\label{appendixmapping}
In this appendix, we will consider the details of the mapping between the backgrounds of supergravity solutions in terms of holomorphic and electrostatic real functions. The backgrounds dual to 5d and 3d SCFTs are examined below. 

\subsection{The DGKU solution dual to 5d SCFTs}

The mapping between the background eq.(\ref{backgroundrescaled}) to the DGKU solution \cite{DHoker:2016ujz} are described in \cite{Legramandi:2021uds}. The DGKU solution parameterises the Riemann surface in the internal space with a complex coordinate $w$, and it is entirely specified by two holomorphic functions $\mathcal{A}_\pm(w)$. 
The metric in string frame
\begin{equation}
d s^2_{10} = e^{\frac{\Phi}{2}}  f_1(w,\bar{w}) \left[ds^2(\text{AdS}_6) + f_2(w,\bar{w}) d s^2 (S^2) + f_3(w,\bar{w}) d w d \bar{w} \right]
\end{equation}
is defined in terms of the following warping functions
\begin{equation}
\label{eq:DGKU_warpings}
f_1 = \frac{|\partial_w \mathcal{G}| \sqrt{1- R^2}}{\kappa \sqrt{R}}  , \quad f_2 = \frac{1}{9} \left(\frac{1-R}{1+R}\right)^2 \, , \quad f_3 = \frac{4 \kappa^4 R}{|\partial_w \mathcal{G}|^2 (1- R^2)^2}\,
\end{equation}
where
\begin{equation}
\label{eq:DGKU_functions}
\begin{split}
&\mathcal{G} = |\mathcal{A}_+|^2-|\mathcal{A}_-|^2+2 \text{Re} \mathcal{B} \, , \qquad \kappa^2 = - \partial_w \partial_{\bar{w}} \mathcal{G} = |\partial_w\mathcal{A}_-|^2-|\partial_w\mathcal{A}_+|^2 \, , \\
&\partial_w \mathcal{B} = \mathcal{A}_+\partial_w\mathcal{A}_- - \mathcal{A}_-\partial_w\mathcal{A}_+ \, , \qquad \, R+R^{-1} = 2 + \frac{6 \kappa^2 \mathcal{G} }{|\partial_w \mathcal{G}|^2} \, ,
\end{split}
\end{equation}
while the fluxes are given by:
\begin{equation}
\label{eq:DGKUfluxes}
\begin{split}
\tau= C_0 + i e^{-\Phi} =& -i \frac{\partial_w(\mathcal{A}_++\mathcal{A}_-) \partial_{\bar{w}} \mathcal{G} - R \partial_{\bar{w}}(\bar{\mathcal{A}}_++\bar{\mathcal{A}}_-) \partial_w \mathcal{G}}{\partial_w(\mathcal{A}_+-\mathcal{A}_-) \partial_{\bar{w}} \mathcal{G} + R \partial_{\bar{w}}(\bar{\mathcal{A}}_+-\bar{\mathcal{A}}_-) \partial_w \mathcal{G}} \, , \\
B_2 +i C_2 =& \frac{2}{3}i\left( \left(\frac{1-R}{1+R}\right)^2 \frac{\partial_w\mathcal{A}_+\partial_{\bar{w}} \mathcal{G}+\partial_{\bar{w}}\bar{\mathcal{A}}_-\partial_w \mathcal{G}}{3 \kappa^2}-\bar{\mathcal{A}}_--\mathcal{A}_+\right) \text{Vol}(S^2) \, .
\end{split}
\end{equation}
The AdS$_6$ radius is set equal to one.

The general form of the functions $\cA_\pm$ was derived in \cite{DHoker:2017mds,Uhlemann:2020bek},
\begin{align}\label{eqn:cA}
 \cA_\pm &=\cA_\pm^0+\sum_{\ell=1}^L Z_\pm^\ell \ln(w-r_\ell)~,
 &
 \overline{\cA_\pm^0}&=-\cA_\mp^0~, & \overline{Z_\pm^\ell}&=-Z_\mp^\ell~.
\end{align}
The poles are on the real line at $r_\ell$, with residues $Z_\pm^\ell$.
These requirements of $\cG$ to be single-valued and positive in the interior of $\Sigma$, and vanishing on the boundary would lead to the regularity conditions
\begin{align}
\label{eqn:constr}
 \cA_+^0 Z_-^k - \cA_-^0 Z_+^k 
+ \sum _{\stackrel{\ell=1}{\ell \neq k} }^LZ^{[\ell, k]} \ln |r_\ell - r_k| &=0~,
&k&=1,\cdots,L~,
\end{align}
where $Z^{[\ell, k]}\equiv Z_+^\ell Z_-^k-Z_+^k Z_-^\ell$.
The residues $Z_\pm^\ell$ encodes the charges of the $(p,q)$ 5-brane emerging at the pole $r_\ell$.

The entire solution is invariant under reparameterization of the complex coordinate $w \to z(w)$. Then, we can use one of the holomorphic functions (or a combination of them) as a definition of the complex coordinate.

\subsubsection{Matching the solutions}

In this section we will show how to match eq.(\ref{backgroundrescaled}) with eq.(\ref{eq:DGKU_warpings})-eq.(\ref{eq:DGKUfluxes}). 
By equating the warping functions $f_1,f_2$ we get the following conditions
\begin{equation}
\label{eq:match1}
\mathcal{G} = \frac{9 \pi^2}{4}\sigma^2 \partial_{\sigma} V_5 \, ,
\qquad \frac{\kappa^2}{|\partial_w \mathcal{G}|^2} = \frac{2}{9 \pi^2 \sigma^2} \frac{\partial _{\eta}^2 V_5}{\Lambda-3\partial _{\eta}^2 V_5\partial _{\sigma }V_5} \, .
\end{equation}
One needs to keep the metric factor in making a comparison for $f_3 d s^2 (\mathbb{C})$. Using the definition of $R$ in eq.(\ref{eq:DGKU_functions}), we have
\begin{equation}
\frac{2}{3} \frac{\kappa^2}{G} d w d \bar{w} = \frac{\partial_{\eta}^2V_5}{3\sigma\partial_{\sigma}V_5} (d \sigma^2 + d \eta^2). 
\end{equation}
Therefore, from equations $\eqref{eq:match1}$ we can write
\begin{equation}
|\partial_w \mathcal{G}|^2 d w d \bar{w} = \left( \left(\partial_\eta \mathcal{G} \right)^2+\left(\partial_\sigma \mathcal{G}\right)^2 \right) (d \sigma^2 + d \eta^2) \, .
\end{equation}
By defining a complex variable $z=\sigma - i \eta$, this consistency relation is automatically solved. Since the DGKU solution is defined up to a change of complex variables, we identify $w=z$ from now on.

Considering the fluxes from  eq.(\ref{backgroundrescaled})
\begin{equation}
B_2 + i C_2 = \frac{2}{3} i \left(\frac{6\pi}{4}V_5-i\frac{3\pi}{4} \eta- i \sigma   \partial_{\sigma} V_5\frac{ \left(-\frac{1}{3}- \frac{2}{3} i \partial_{\eta} V_5 \right) \partial_{\sigma \eta}^2 V_5+2 i \partial_{\sigma} V_5 \partial_{\eta}^2 V_5 }{\Lambda}\right)\text{Vol}(S^2) \, 
\end{equation}
and comparing it with  eq.(\ref{eq:DGKUfluxes}), the two expressions match if we set
\begin{equation}
\label{Apm_eq}
\mathcal{A}_++\bar{\mathcal{A}}_- = i\frac{3\pi}{4} \eta-\frac{3\pi}{2} \partial_\sigma (\sigma V_5) \, .
\end{equation}
Where, $\sigma V_5 = \hat{V_5}$ is the harmonic function defined in  eq.(\ref{eqfinal}) and, since it is also real, it defines just one holomorphic function $\mathcal{V}(z)$
\begin{equation}
\sigma V_5 = \mathcal{V}(z) + \overline{\mathcal{V}(z)}.
\end{equation}
Using this condition and the fact that $\mathcal{A}_+$ is holomorphic while $\bar{\mathcal{A}}_-$ is anti-holomorphic, we see that  eq.(\ref{Apm_eq}) completely defines $\mathcal{A}_\pm$ in terms of $\sigma V_5$ and the new coordinate $z$
\begin{equation}
\label{eq:A1A2def}
\mathcal{A}_\pm = \mp \frac{3\pi z}{8} - \frac{3\pi}{2}\partial_z (\sigma V_5) \, .
\end{equation}
With these definitions, the reader can check that the axion-dilaton expressions are identical.

In order to compare the two backgrounds, we had to impose $w=z$. As a consequence, the two holomorphic functions $\mathcal{A}_\pm$ are defined just in terms of one of the holomorphic function ($\mathcal{V}$) and the coordinate $z$. 

It should be noted that one can do the rescaling $V_5 = \nu V_{5, \text{old}}$ and $(\sigma,\eta)= (\mu  \sigma_{\text{old}},\mu  \eta_{\text{old}})$ with corresponding changes in the background to make quantised Page charges of the branes integers. In \cite{Legramandi:2021uds}, this rescaling is done to quantise charges properly. In the case of holomorphic functions, this can also be done to get integer charges which match with the ones obtained from the Rank function in the real formalism. One can also perform the inverse of that rescaling again on our potential and coordinates to match the results. From now on we do the change $(\sigma,\eta) \rightarrow 2 (\sigma,\eta)$ and $V_5 \rightarrow V_5$ to compare results.   

\subsubsection{Example: $T_P$ theory} \label{subsec:Tn_Uhul}
In this section, we consider solutions with three poles, mapped in terms of our potential $V_5$. We refer to \cite{Uhlemann:2019ypp} for the specific form of the solutions we are considering.

The $T_P$ theory is given by the three-pole solution
\begin{equation}
\mathcal{A}_\pm = \frac{3 P}{4} (\pm \log (w-1)+(\mp 1-i) \log (w+1)+i \log (2 w)) \, ,
\end{equation}
where the poles are at $w=1,0,-1$ and we set $\alpha'=1$. The coordinate $z = \sigma - i \eta$ is defined as
\begin{equation}
z = -\frac{2}{3 \pi } (\mathcal{A}_+-\mathcal{A}_-) = \frac{N}{\pi} \log \left( \frac{1+w}{1-w} \right) \quad \Rightarrow \quad w = \coth \left(\frac{\pi  z}{2P}\right) \, .
\end{equation}
Notice that the imaginary axis for $w$ becomes the interval $\eta \in (0,  N)$ at $\sigma = 0$, while the real axis, which is the space-time boundary, is mapped as following
\begin{equation}
\begin{split}
w \in (-1,1) \quad &\Rightarrow \quad \eta = P \, , \, \sigma \in (- \infty, \infty) \, , \\ \nonumber
w \in (-\infty,-1) \cup (1,\infty) \quad &\Rightarrow \quad \eta = 0 \, , \, \quad \sigma \in (- \infty, \infty) \, .
\end{split}
\end{equation}
So the space-time boundary in the $w$ coordinate is consistently mapped in the space-time boundary in $\sigma$ and $\eta$ coordinates.

The potential is defined by the following equation
\begin{equation}
\partial_z (\sigma V_5) = -\frac{\mathcal{A}_- + \mathcal{A}_+}{3\pi} = -\frac{i P}{2 \pi } \log \left(e^{-\frac{ \pi z}{ P}}+1\right)
\end{equation}
which can be integrated leading to
\begin{align}
V_5 =& \frac{ i P^2 }{2 \pi ^2 \sigma } \left(\text{Li}_2\left(-e^{-\frac{\pi  (|\sigma|+ i\eta )}{P}}\right)-\text{Li}_2\left(-e^{\frac{- \pi  (|\sigma|- i\eta )}{ P}}\right)\right) \nonumber \\[2mm]
=& \frac{P^2 }{\pi^2 \sigma}\sum _{k=1}^{\infty }  \frac{(-1)^{k+1}}{k^2} \sin \left(\frac{k \pi}{ P} \eta \right) e^{-\frac{k \pi}{P}|\sigma|} \label{eq:T_N-U} \, ;
\end{align}
the integration constant is set to zero as required by the boundary conditions.
eq.(\ref{eq:T_N-U}) is exactly of the form eq.(\ref{eq:fourier_vhat}), and we can identify the coefficient of the Fourier expansion:
\begin{equation}
a_k = \frac{ P^2 }{\pi^2} \frac{(-1)^{k+1}}{k^2} \, .
\end{equation}
This result can be compared with eq.(\ref{eq:Trank}) for $N=1$, which exactly matches.
\subsubsection{Wilson loops}
Interestingly, the Wilson loop expectation value in $\mathbb{k}^{th}$ anti symmetric representation is proportional to $\cG$ \cite{Uhlemann:2020bek} which in the real formalism from eq.(\ref{eq:match1}) is simply proportional to $\sigma^2 \partial_{\sigma} V_5$, evaluated at a point determined by gauge node and anti symmetric representation chosen
\begin{equation}
    \ln\langle W_\wedge\rangle =-\frac{2}{3}T_{\rm D3} {\rm Vol}_{AdS_2}{\rm Vol}_{S^2} \cG=3\pi\sigma^2\left.\partial_\sigma V_5\right|_{(\sigma*,\eta*)},
\end{equation}
with $T_{\rm D3} {\rm Vol}_{AdS_2}{\rm Vol}_{S^2}=-4\pi/(2\pi \alpha^\prime)^2$. So the Wilson loop is
\begin{equation}
    \ln\langle W_\wedge\rangle =3\pi \left.(\sigma \partial_{\sigma} \hat V -\hat V )\right|_{(\sigma*,\eta*)}=3\pi 
    \sum_{k=1}^\infty \frac{\mathcal{R}_k}{2}(\frac{P}{k\pi}) \sin\left(\frac{k\pi}{P}\eta^* \right) {e^{-\frac{k\pi}{P}|\sigma^*|}}(\frac{k\pi}{P}|\sigma^*|+1).
\end{equation}
The only calculation needed is the location on the $\Sigma$-plane, which this function should be calculated. So
\begin{equation}
    N_{\rm F1}+i N_{\rm D1}=\frac{2}{3\pi}\Big[ \frac{i}{3}(\frac{9\pi}{2}) \eta^*-6 (\frac{\pi}{4})\left.\partial_\sigma (\hat V)\right|_{(\sigma*,\eta*)}\Big]
\end{equation}
$N_{\rm F1}$ and $N_{\rm D1}$ are related to the gauge node and anti symmetric representation chosen. Hence,
\begin{equation}
    N_{\rm D1}= \eta^*, \quad N_{\rm F1}=\sum_{k=1}^\infty \frac{\mathcal{R}_k}{2} \sin\left(\frac{k\pi}{P}\eta^* \right) {e^{-\frac{k\pi}{P}|\sigma^*|}}\operatorname{Sgn}(\sigma^*).
\end{equation}

\subsection{The DEGK background dual to 3d SCFTs}\label{mapDEGK}
The DEGK solutions \cite{DHoker:2007hhe} in the string frame is  defined in terms of complex functions in the variable $w$:
\begin{eqnarray}
& & ds_{10,st}^2= f_1(w, \bar{w})\Big[ds^2(\text{AdS}_4) + f_2(w,\bar{w}) d s^2 (S^2_1)+ f_3(w,\bar{w}) d s^2 (S^2_2)+ f_4(w,\bar{w})d w d \bar{w} \Big], \nonumber\\[2mm]
& &e^{-2\Phi}=f_5(w,\bar{w}), \;\;\;\; B_2=f_6(w,\bar{w}) \text{Vol}(S^2_1),\;\;\;\; C_2= f_7(w,\bar{w}) \text{Vol}(S^2_2),
\end{eqnarray}
where
\begin{eqnarray}
& & f_1 = 2 \sqrt{-\frac{N_2 }{W}}, \;\;\;\; f_2 = -\frac{h_1^2 W}{N_1}, \;\;\;\; f_3 = -\frac{h_2^2 W}{N_2},\;\;\;\; f_4 = -2\frac{W}{h_1 h_2} , \;\;\;\; f_5 = \frac{N_1}{N_2} \nonumber\\[2mm]
& & f_6 = 4 \frac{h_1^2 h_2 \text{Im}(\partial_w h_2 \partial_{\bar{w}} h_1)}{N_1}+2 h_2^D , \;\;\;\; f_7 = 4 \frac{h_1 h_2^2 \text{Im}(\partial_w h_2 \partial_{\bar{w}} h_1)}{N_2}-2 h_1^D . \label{eq:def-fi_holomorphic}
\end{eqnarray}
The five-form field is given by
\begin{equation}
F_5 = \text{Vol(AdS}_4) \wedge d f_8+ *( \text{Vol(AdS}_4) \wedge d f_8),
\end{equation}
and
\begin{equation}
f_8 = 4 \left( 6 \text{Re} (\mathcal{C})- 3 \mathcal{D} - 2\frac{h_1 h_2}{W} \text{Im}(\partial_w h_1 \partial_w h_2) \right) \, .
\end{equation}
All these functions could be defined in terms of two holomorphic functions $\mathcal{A}_{1,2}(w)$, in particular $h_{1,2}$ and $h_{1,2}^D$ are the dual real harmonic functions
\begin{equation}
h_1 = 2 \text{Im} (\mathcal{A}_1) \, , \qquad h_1^D = 2 \text{Re} (\mathcal{A}_1) \, , \qquad h_2 = 2 \text{Re} (\mathcal{A}_2) \, , \qquad h_2^D = -2 \text{Im} (\mathcal{A}_2).
\end{equation}
Also, we have  the following definitions
\begin{equation}
W= \partial_w\partial_{\bar{w}} (h_1 h_2) \, , \quad N_i = 2 h_1 h_i |\partial_w h_i|^2-h_i^2 W \, , \quad \mathcal{D} = 2 \text{Re}(\mathcal{A}_1 \bar{\mathcal{A}}_2) \, , \quad \partial_w \mathcal{C} = \mathcal{A}_1 \partial_w \mathcal{A}_2 - \mathcal{A}_2 \partial_w \mathcal{A}_1 .
\end{equation}
A more detailed background solution description could be found in \cite{Coccia:2021lpp}.

Solutions with different holographic interpretations can be constructed depending on the choice of $h_{1 / 2}$ and $\Sigma$. We would be interested in duals of $3\mathrm{~d}$ SCFTs. All solutions here describe D3-branes suspended between, ending on, or intersecting combinations of D5 and NS5 branes. For these solutions, the harmonic functions $h_{1}, h_{2}$ on the strip
$$
\Sigma=\left\{w \in \mathbb{C} \mid 0 \leq \operatorname{Im}(w) \leq \frac{\pi}{2}\right\}
$$
would read
$$
\begin{aligned}
h_{1} &=-\frac{\alpha^{\prime}}{4} \sum_{a=1}^{A} N_{\mathrm{D} 5}^{(a)} \ln \tanh \left(\frac{i \pi}{4}-\frac{w-\delta_{a}}{2}\right)+\text { c.c. } \\
h_{2} &=-\frac{\alpha^{\prime}}{4} \sum_{b=1}^{B} N_{\mathrm{NS} 5}^{(b)} \ln \tanh \left(\frac{w-\delta_{b}}{2}\right)+\text { c.c. }.
\end{aligned}
$$
These solutions describe $A$-groups of D5-branes with $N_{\text {D5 }}^{(a)}$ D5-branes in the $a^{\text {th }}$ group and $B$-groups of NS5-branes with $N_{\text {NS5 }}^{(b)}$ NS5-branes in the $b^{\text {th }}$ group. D3-branes are suspended between the 5 -branes for 3d SCFTs.

The background is invariant under conformal transformations $w \to f(w)\equiv z$; specifically, one can choose one of the holomorphic functions as a coordinate. The second holomorphic function can be defined in terms of an auxiliary harmonic function $\hat{V}_3(z,\bar{z})$ as follows
\begin{equation}
\mathcal{A}_1 = \pi \partial_z \hat{V}_3 \, , \qquad \mathcal{A}_2= \frac{\pi}{8} z \, .\label{B7}
\end{equation}
In order to match these backgrounds with those in eq.(\eqref{background}), we can set
\begin{equation}
z = \sigma - i \eta \, , \qquad \hat{V}_3 = \sigma V_3 \, .
\end{equation}
With these identifications,
\begin{eqnarray}
& & h_1 = \pi \sigma \partial_\eta V_3 \, , \qquad h_1^D = \pi \partial_\sigma (\sigma V_3) \, , \qquad h_2 = \frac{\pi}{4} \sigma \, , \qquad h_2^D = \frac{\pi}{4} \eta \, , \\[2mm]
& & W= \frac{\pi^2}{8} \partial_{\sigma}(\sigma \partial_\eta V_3) \, , \qquad N_1 = \frac{\pi^4}{8}\sigma^3 \partial_\eta V_3 \Lambda \, , \qquad N_2 = - \frac{\pi^4}{128} \sigma^3 \partial^2_{\eta\sigma} V_3 \, . 
\end{eqnarray}
These expressions will match eq.(\ref{eq:def-fi_holomorphic}) with eq.(\ref{background}). 
For the same reason as the 5d case, we do the change $(\sigma,\eta) \rightarrow 2 (\sigma,\eta)$ and $V_3 \rightarrow  V_3$ to quantize the Page charges properly and match the backgrounds.   

\subsubsection{Example: Generic balanced quivers} \label{subsec:BalancedQ}
The backgrounds dual to generic balanced quivers are given by
\begin{align}
h_{1} &=-\frac{\alpha^{\prime}}{4} \sum_{a=1}^{A} N_{\mathrm{D} 5}^{(a)} \ln \tanh \left(\frac{i \pi}{4}-\frac{w-\delta_{a}}{2}\right)+\text { c.c. } \\ \nonumber
h_{2} &=-\frac{\alpha^{\prime}}{4} N_{\mathrm{NS} 5} \ln \tanh \left(\frac{w}{2}\right)+\text { c.c. }.
\end{align}
The dual is a quiver with $N_{\mathrm{NS} 5}-1$ nodes and $N_{\text {D5 }}^{(a)}$ flavors at gauge nodes $\mathrm{t}_{a}$ with
$$
\mathrm{t}_{a}=\frac{2}{\pi} N_{\text {NS5 }} \arctan e^{\delta_{a}}
$$
Since all nodes are balanced and $N_{0}=N_{L+1}=0$, the entire quiver can be reconstructed from this information.
These functions can be written in terms of 
\begin{align}
\mathcal{A}_1  &=-i\frac{\alpha^{\prime}}{4} \sum_{a=1}^{A} N_{\mathrm{D} 5}^{(a)} \ln \tanh \left(\frac{i \pi}{4}-\frac{w-\delta_{a}}{2}\right) \\ \nonumber
\mathcal{A}_2  &=-\frac{\alpha^{\prime}}{4} N_{\mathrm{NS} 5} \ln \tanh \left(\frac{w}{2}\right).
\end{align}
It must be noted that one can multiply the argument of logarithm in $\mathcal{A}_1$ and $\mathcal{A}_2$functions with a phase $|c|=1$ (for instance, $\mathcal{A}_2\rightarrow -\frac{ \alpha^{\prime}}{4} N_{\mathrm{D} 5} \ln c \tanh \left(-\frac{w}{2}\right)$ ), while $h_{1,2}$ functions remain the same. 

If we take $z$ to be the new coordinate, our transformation will read (set $\alpha'=1$)
\begin{equation}
	z=-\frac{1}{\pi} N_{\mathrm{NS} 5} \ln \tanh \left(\frac{w}{2}\right).
\end{equation}
This change of coordinates could be done also in the following steps. First, $w'=e^w$ sends the strip to the upper right quadrant with NS5 on reals and D5s on the imaginary axis. Second, $\frac{w'-1}{w'+1}=-u \quad  u' = -u$ sending the upper right quadrant to the upper half disk with NS5 on zero and D5s on the circumference. Lastly, $z = -\frac{1}{\pi} N_{\mathrm{NS} 5} \ln (u)$ which maps upper half disk to an strip $0 \leq \operatorname{Re}(z') < \infty$ and $0 \leq \operatorname{Im}(z') < -N_{NS5}$. The Neveu-Schwarz five barnes are mapped to a vertical line at infinity, and D5s are on $\operatorname{Re}(z')=0$. By taking $z=\sigma-i \eta$ and the condition $V(-\sigma,\eta)=-V(\sigma,\eta)$ the solution would be well defined on $0 \leq \eta \leq + N_{\mathrm{NS} 5}$.

After the mapping, the holomorphic functions are
\begin{align}
\mathcal{A}_{1}(z) &=-\frac{i}{4} \sum_{a} N_{\mathrm{D} 5}^{(a)}\left[\ln \left(1-e^{\frac{-\pi}{2\alpha'N_{\mathrm{NS} 5}}z'} / \sigma_{a}\right)-\ln \left(1-\sigma_{a} e^{\frac{-\pi}{2\alpha'N_{\mathrm{NS} 5}}z'}\right)\right] \equiv \pi \partial_{z} \widehat{V}_3 \\
\mathcal{A}_{2}(z) &=\frac{\pi}{4} z ,
\end{align}
with $\sigma_{a}=\frac{i e^{\delta_{a}}-1}{i e^{\delta_{a}}+1}=e^{\frac{-i\pi t_a}{N_{\mathrm{NS} 5}}}$.
Then
\begin{equation}
	\int \mathcal{A}_{1}(z) \operatorname{d}z = \frac{i}{4} (\frac{-N_{\mathrm{NS} 5}}{\pi})\sum_{a} N_{\mathrm{D} 5}^{(a)}\left[\operatorname{Li_2} \left(\sigma_{a} e^{\frac{-\pi}{N_{\mathrm{NS} 5}}z}\right)-\operatorname{Li_2} \left( e^{\frac{-\pi}{N_{\mathrm{NS} 5}}z}/\sigma_{a}\right)\right] \equiv f(z),
\end{equation}
hence $\widehat{V}_3 = \frac{1}{\pi} f(z)+c.c.$. The integration constant is chosen to make the $\widehat{V}_3$ function harmonic and the constants $c_a=1/ \sigma_a$ are chosen to keep the boundary conditions after mapping.
Then,
\begin{equation} \label{eq:genericbq3}
	 \widehat{V}_3= (\frac{N_{\mathrm{NS} 5}}{2\pi^2})\sum_{a} N_{\mathrm{D} 5}^{(a)}\operatorname{Im}\left[\operatorname{Li_2} \left(\sigma_{a} e^{\frac{-\pi}{N_{\mathrm{NS} 5}}z}\right)-\operatorname{Li_2} \left( e^{\frac{-\pi}{N_{\mathrm{NS} 5}}z}/\sigma_{a}\right)\right] , 
\end{equation}
which matches with the last line of eq.(\ref{potentialhat3}) with identification $N_{\mathrm{NS} 5}\equiv P$, $N_{\mathrm{D} 5}^{(a)}\equiv F_J$ and $t_a\equiv J$.

\subsubsection{Wilson loops}
With $\hat V_3^D$ dual to $\hat V_3$ and $\partial_{\eta} \hat W = \hat V_3$ the wilson loop computed in \cite{Coccia:2021lpp} would read
\begin{align}
	\ln \left\langle W_{\wedge}\right\rangle & =\frac{8}{\pi^{2} \alpha^{\prime 3}}\int d \xi h_{1} h_{2}\left(\partial_{z} h_{2}\right) z^{\prime}
= \pi \left[ \int \sigma \partial_\eta \hat V_3(\sigma,\eta) \mathrm{d}\sigma\right]_{(\infty,\eta*)}^{(\sigma*,\eta*)} \\ \nonumber
& =\pi \left[ \sigma \int  \partial_\eta \hat V_3(\sigma,\eta) \mathrm{d}\sigma-\int \int \partial_\eta \hat V_3 \mathrm{d}\sigma\mathrm{d}\sigma'\right]_{(\infty,\eta*)}^{(\sigma*,\eta*)} = \pi \left[-\sigma \partial_\sigma \hat W(\sigma,\eta) + \hat W(\sigma,\eta)\right]_{(\infty,\eta*)}^{(\sigma*,\eta*)}.
\end{align}
A by part integration is done in the second line and one can check that $\partial_\sigma \hat W(\sigma,\eta)=-\int  \partial_\eta \hat V_3(\sigma,\eta) \mathrm{d}\sigma$. The limits can be computed easily from the rank and node of the Wilson loop in
\begin{equation}
N_{\mathrm{F} 1}=\frac{4}{\pi^{2} \alpha^{\prime 2}}\left[\operatorname{Im}\left(\mathcal{A}_{1} \mathcal{A}_{2}+\mathcal{C}\right)\right]_{(\infty,\eta*)}^{(\sigma*,\eta*)}=\left[\hat V_3 ^D\right]_{(\infty,\eta*)}^{(\sigma*,\eta*)}, \quad N_{\mathrm{D} 3}=\frac{4}{\pi \alpha'} h_{2}^{D}\equiv \eta^* .
\end{equation}
$\alpha'=1$ is chosen as above. It can be deduced from the second equation that the integration limits are along constant $\eta$. Indeed, the rank and node of the Wilson loop determine the $N_{\mathrm{F} 1}$ and $N_{\mathrm{D} 3}$ on the probe D5 brane, which should be embedded in the background geometry to calculate the expectation value.  Hence, its trajectory on the $\Sigma$ surface is calculated to satisfy the BPS conditions. ${z_{0}}$ and ${z_{1}}$ are endpoints of this trajectory, and integrals are along this curve. More details can be found in the given references.

In terms of the Fourier expansion of potential fields with definitions
\begin{equation}
\hat{V}(\sigma,\eta)= \begin{cases}
\sum_{k=1}^\infty  \frac{a_k}{2} \left( e^{-\frac{k\pi}{P}z} + e^{-\frac{k \pi}{{P}} \bar{z} }\right)=\operatorname{Re}\sum_{k=1}^\infty  a_k \left( e^{-\frac{k\pi}{P}z}\right)  & \sigma \ge 0, \\[2mm]
\sum_{k=1}^\infty  \frac{a_k}{2} \left( e^{\frac{k\pi}{P} z} + e^{\frac{k \pi}{{P}} \bar{z} }\right)=\operatorname{Re}\sum_{k=1}^\infty a_k \left( e^{\frac{k\pi}{P}z}\right) & \sigma < 0 ,
\end{cases}
\end{equation}
and for $\hat V ^D$ we have ($\operatorname{Re} \mapsto \operatorname{Im}$)
\begin{equation}
\hat{V}^D(\sigma,\eta)= \begin{cases}
\sum_{k=1}^\infty  a_k \sin\left( \frac{k\pi\eta}{P}\right) e^{-\frac{k \pi \sigma}{P}} & \sigma \ge 0, \\[2mm]
-\sum_{k=1}^\infty  a_k \sin\left( \frac{k\pi\eta}{P}\right) e^{\frac{k \pi \sigma}{P}} & \sigma < 0 .
\end{cases}
\end{equation}
Hence the Wilson loop is
\begin{equation}
    \ln\langle W_\wedge\rangle =
     \pi\sum_{k=1}^\infty \frac{\mathcal{R}_k}{2}(\frac{P}{k\pi}) \sin\left(\frac{k\pi}{P}\eta^* \right) {e^{-\frac{k\pi}{P}|\sigma^*|}}(\frac{k\pi}{P}|\sigma^*|+1).
\end{equation}
The evaluation point is
\begin{equation}
    N_{D3}=\eta^*, \quad N_{\mathrm{F} 1}=\hat V ^D=\sum_{k=1}^\infty \frac{\mathcal{R}_k}{2} \sin\left(\frac{k\pi}{P}\eta^* \right) {e^{-\frac{k\pi}{P}|\sigma^*|}}\operatorname{Sgn}(\sigma^*).
\end{equation}

\subsubsection{An example of mapping S-dual backgrounds}\label{appS}
In this section, we consider mappings of the triangular quivers considered in section \ref{mirrorsection}. For the triangular quiver, one has
$$
\begin{aligned}
h_{1} &=-\frac{i}{4} N_{\mathrm{D} 5} \ln \left|\tanh \left(\frac{i \pi}{4}-\frac{w-\delta}{2}\right)\right| ,\\
h_{2} &=-\frac{1}{4} N_{\mathrm{NS} 5} \ln\left| \tanh \frac{w}{2}\right| ,\\
\mathcal{A}_{1} &=-\frac{i}{4} N_{\mathrm{D} 5} \ln \tanh \left(\frac{i \pi}{4}-\frac{w-\delta}{2}\right) ,\\
\mathcal{A}_{2} &=-\frac{1}{4} N_{\mathrm{NS} 5} \ln \tanh \left(\frac{w}{2}\right).
\end{aligned}
$$ 
Our mapping would be
\begin{equation}
	z=-\frac{1}{\pi} N_{\mathrm{NS} 5} \ln \tanh \left(\frac{w}{2}\right),
\end{equation}
giving
\begin{align}
\mathcal{A}_{1}(z') &=-\frac{i \alpha^{\prime}}{4} \sum_{a} N_{\mathrm{D} 5}^{(a)}\left[\ln \left(1-e^{\frac{-\pi}{2\alpha'N_{\mathrm{NS} 5}}z'} / \tilde \sigma_{a}\right)-\ln \left(1-\tilde\sigma_{a} e^{\frac{-\pi}{2\alpha'N_{\mathrm{NS} 5}}z'}\right)\right] \equiv \pi \partial_{z'} \widehat{V} ,\\
\mathcal{A}_{2}(z') &=\frac{\pi}{8} z',
\end{align}
with $\sigma=\frac{i e^{\delta}-1}{ie^{\delta}+1} =e^{\frac{-i\pi t}{N_{\mathrm{NS} 5}}}$ and $t$ is the gauge node for which the flavour is inserted. Here the constant $c=1/\tilde \sigma$ is chosen to keep the boundary conditions after mapping.
Then from eq.(\ref{eq:genericbq3}) one finds
\begin{equation} 
	 \widehat{V}^{elec}_3= (\frac{N_{\mathrm{NS} 5}}{2\pi^2}) N_{\mathrm{D} 5}\operatorname{Im}\left[\operatorname{Li_2} \left(\sigma e^{\frac{-\pi}{N_{\mathrm{NS} 5}}z}\right)-\operatorname{Li_2} \left( e^{\frac{-\pi}{N_{\mathrm{NS} 5}}z}/\sigma\right)\right] .
\label{ccxx}
\end{equation}

The S-dual configuration can be obtained from exchanging $h_1$ and $h_2$
$$
\begin{aligned}
\mathcal{A}_{1} &=-\frac{i}{4} N_{\mathrm{NS} 5} \ln \tanh \left(\frac{w}{2}\right) \, ,\\
\mathcal{A}_{2} &= -\frac{1}{4} N_{\mathrm{D} 5} \ln \tanh \left(\frac{i \pi}{4}-\frac{w-\delta}{2}\right) .
\end{aligned}
$$
The desired transformation is
\begin{equation}
	z=-\frac{2\alpha^{\prime}}{\pi} N_{\mathrm{NS} 5} \ln \tanh \left(\frac{i \pi}{4}-\frac{w-\delta}{2}\right).
\end{equation}
The transformation can be done in two steps
\begin{equation}
	\frac{z'}{2}=\frac{i \pi}{4}-\frac{w-\delta}{2}, \quad z=-\frac{1}{\pi} N_{\mathrm{NS} 5} \ln \tanh \left(\frac{z'}{2}\right).
\end{equation}
After the first step, the holomorphic functions would be
$$
\begin{aligned}
\mathcal{A}_{1} &=-\frac{i}{4} N_{\mathrm{NS} 5} \ln \tanh \left(\frac{i \pi}{4}-\frac{z'-\delta}{2}\right),  \\
\mathcal{A}_{2} &=-\frac{1}{4} N_{\mathrm{D} 5} \ln \tanh \left(\frac{z'}{2}\right),
\end{aligned}
$$
which is exactly the one before S-duality with $N_{\mathrm{NS} 5} \longleftrightarrow N_{\mathrm{D} 5}$. The range of $\eta$ under final transformation would be $(0,N_{\mathrm{D} 5})$ instead of $(0,N_{\mathrm{NS} 5})$ but the gauge node insertion relative to the range would be the same as before S-duality,
\begin{equation} 
	 \widehat{V}^{mag}_3= (\frac{N_{\mathrm{D} 5}}{2\pi^2}) N_{\mathrm{NS} 5}\operatorname{Im}\left[\operatorname{Li_2} \left(\sigma e^{\frac{-\pi}{N_{\mathrm{D} 5}}z}\right)-\operatorname{Li_2} \left( e^{\frac{-\pi}{N_{\mathrm{D} 5}}z}/\sigma\right)\right] .
\label{xxc}\end{equation}

The relation between eqs.(\ref{ccxx}) and (\ref{xxc}) can otherwise be obtained by applying the generic expression for the potential $\widehat{V}_3$ in eq.(\ref{potentialhat3}), in the case of no-offsets,  and using eqs.(\ref{relacionesEM}). This makes the point that the electrostatic version of mirror symmetry, encodes S-duality.

\end{document}